\pdfoutput=1
\documentclass[12pt,a4paper]{article}
\usepackage[utf8]{inputenc}
\usepackage{jheppub}
\hypersetup{unicode=true,bookmarksopen=true}

\usepackage{mathtools}
\usepackage{lmodern}
\usepackage[T1]{fontenc}
\usepackage{bbm}
\DeclareMathAlphabet{\mathbfi}{OML}{cmm}{b}{it}

\let\originalleft\left
\let\originalright\right
\renewcommand{\left}{\mathopen{}\mathclose\bgroup\originalleft}
\renewcommand{\right}{\aftergroup\egroup\originalright}

\makeatletter
\newenvironment{equations}[1][]{\subequations\ifx\relax#1\relax\else\label{#1}\fi\align\ignorespaces}{\endalign\ignorespacesafterend\endsubequations}
\def\@spliteq#1{\begin{equation}\begin{split}#1\end{split}\end{equation}}
\def\splitequation{\collect@body\@spliteq}

\makeatother

\renewcommand{\vec}[1]{{\ifnum9<1#1\mathbf{#1}\else\ifcat\noexpand#1\relax\boldsymbol{#1}\else\mathbfi{#1}\fi\fi}}
\newcommand{\mathe}{\mathrm{e}}
\newcommand{\mathi}{\mathrm{i}}
\let\oldre\Re
\let\oldim\Im
\renewcommand{\Re}{\oldre\mathfrak{e}\,}
\renewcommand{\Im}{\oldim\mathfrak{m}\,}
\newcommand{\total}{\mathop{}\!\mathrm{d}}

\newcommand{\1}{\mathbbm{1}}
\newcommand{\tr}{\operatorname{tr}}
\newcommand{\eqend}[1]{\,#1}
\newcommand{\bigo}[1]{\mathcal{O}\left({#1}\right)}

\makeatletter
\def\expect{\@ifnextchar[{\@expecttw@}{\@expect@ne}}
\def\@expecttw@[#1]#2{\left\langle{#2}\right\rangle^\text{#1}}
\def\@expect@ne#1{\left\langle{#1}\right\rangle}
\makeatother
\newcommand{\dn}[2][n]{\frac{\total^{#1} {#2}}{(2\pi)^{#1}}}

\newcommand{\trlss}[1]{\left[ #1 \right]_{\tr}}
\newcommand{\MS}{\ensuremath{\overline{\text{MS}}}\ }
\newcommand{\diver}{\mathcal{N}}

{%
\end{oldthebibliography}%
}

\allowdisplaybreaks

\title{Trace anomaly for Weyl fermions using the Breitenlohner--Maison scheme for $\gamma_*$}

\author[a]{S.~Abdallah,}
\author[b,c]{S.~A.~Franchino-Vi{\~n}as}
\author[a]{and M.~B.~Fr{\"o}b}

\affiliation[a]{Institut f{\"u}r Theoretische Physik, Universit{\"a}t Leipzig,\\ Br{\"u}derstra{\ss}e 16, 04103 Leipzig, Germany}
\affiliation[b]{Departamento de F{\'\i}sica, Facultad de Ciencias Exactas, Universidad Nacional de\\ La Plata, C.C.\ 67 (1900), La Plata, Argentina}
\affiliation[c]{Institut f{\"u}r Theoretische Physik, Universit{\"a}t Heidelberg, 69120 Heidelberg, Germany}

\emailAdd{sami.abdallah@itp.uni-leipzig.de}
\emailAdd{safranchino@fisica.unlp.edu.ar}
\emailAdd{mfroeb@itp.uni-leipzig.de}

\abstract{We revisit the computation of the trace anomaly for Weyl fermions using dimensional regularization. For a consistent treatment of the chiral gamma matrix $\gamma_*$ in dimensional regularization, we work in $n$ dimensions from the very beginning and use the Breitenlohner--Maison scheme to define $\gamma_*$. We show that the parity-odd contribution to the trace anomaly vanishes (for which the use of dimension-dependent identities is crucial), and that the parity-even contribution is half the one of a Dirac fermion. To arrive at this result, we compute the full renormalized expectation value of the fermion stress tensor to second order in perturbations around Minkowski spacetime, and also show that it is conserved.}

\begin{document}

\maketitle

\section{Introduction}
\label{sec:introduction}

In a series of recent articles it has been debated whether there could be unitarity issues at the one-loop level in theories involving chiral fermions in four-dimensional curved spacetimes~\cite{Duff:2020dqb}. Elaborating on some previous results~\cite{Christensen:1978md,Nakayama:2012gu}, the group of Bonora et al. has claimed that there exists a CP-odd term in the trace anomaly of Weyl fermions, whose coefficient is purely imaginary~\cite{Bonora:2014qla,Bonora:2015nqa,Bonora:2017gzz}. These results were derived in dimensional regularization, both using a standard perturbative computation around Minkowski spacetime using Feynman diagrams and introducing an ``metric-axial-tensor gravity'' approach similar to the axial vector potential in gauge theory. However, with strictly four-dimensional regularization methods such as Pauli--Villars regularization~\cite{Bastianelli:2016nuf,Bastianelli:2019zrq} employed by Bastianelli et al., or Hadamard subtraction used by Zahn and the last author~\cite{Frob:2019dgf}, CP-odd terms in the trace anomaly do not arise at all. In between, this has given rise to more general discussions involving gauge anomalies~\cite{Bastianelli:2018osv,Bastianelli:2019fot} and general CP-violating theories~\cite{Nakayama:2018dig,Nakagawa:2020gqc}. Apart from the natural consequences this may have in the building of the Standard Model and extensions thereof, there will also be repercussions in our understanding of the early universe~\cite{Dolgov,Fabris:1998vq,Hawking:2000bb}. Other recent related results concern the anomalies for non-relativistic fermions~\cite{Auzzi:2017jry} and the anomalies in the Standard-Model Effective Field Theory~\cite{Cata:2020crs,Feruglio:2020kfq}.

In~\cite{Bonora:2014qla,Bonora:2015nqa,Bonora:2017gzz} the authors have made use of dimensional regularization~\cite{Bollini:1972ui,tHooft:1972tcz,Leibbrandt:1975dj}. As is well-known since its birth~\cite{tHooft:1972tcz,Bollini:1973wu}, the definition of the chiral $\gamma_*$ matrix in this framework is subtle, and many proposals have been made; see~\cite{Baikov:1991qz} for a comparison. It seems to us that the following prescriptions are the most promising solutions:
\begin{itemize}
\item The Breitenlohner--Maison scheme~\cite{Breitenlohner:1977hr}, which is an improvement of the original proposal made in~\cite{tHooft:1972tcz}. It formally breaks $n$-dimensional Lorentz covariance by writing the full spacetime as a direct product, one piece being strictly four-dimensional and the other having $(n-4)$ dimensions. The chiral $\gamma_*$ anticommutes with the first four $\gamma$ matrices, while it commutes with the other $n-4$ ones. However, cyclicity of the matrix trace is preserved as well as the relation $\gamma_*^2 = \1$; we will expand on this in Sec.~\ref{sec:BMprescription}.
\item The Thompson--Yu prescription~\cite{Thompson:1985uv}, where a non-vanishing anticommutator of $\gamma_*$ with all other $\gamma$ matrices is introduced. This maintains both the cyclicity of the trace and $\gamma_*^2 = \1$ as well as Lorentz invariance. However, the expressions are much lengthier than in the previous case and computations become rather cumbersome.
\item The Kreimer et al.\ prescription~\cite{Korner:1991sx,Kreimer:1989ke}, where one can keep the vanishing anti-commutator of $\gamma_*$ with all other $\gamma$ matrices. However, this requires the embedding of the four-dimensional $\gamma$-matrix algebra in an infinite-dimensional one, with the loss of cyclicity of the trace and a complicated prescription for the evaluation of traces containing $\gamma_*$ as a consequence.
\end{itemize}
We note that from this list, only the Breitenlohner--Maison scheme has been shown to give mathematically consistent results to arbitrary loop orders~\cite{Breitenlohner:1977hr,Rufa:1990hg}. In this work we are thus going to analyze the trace anomaly for a chiral fermion strictly following this scheme in dimensional regularization. Apart from this point and the fact that we work consistently in $n$ dimensions from the beginning, our method is completely analogous to the one of Bonora et al.~\cite{Bonora:2014qla,Bonora:2015nqa}, i.e., an expansion to second order in perturbations around Minkowski spacetime.

The article is organized as follows: In Sec.~\ref{sec:weyl} we review the definition of a Weyl fermion in curved space. In particular, we include in Sec.~\ref{sec:BMprescription} a description of the regularization scheme that will be employed, while in Sec.~\ref{sec:expansion} we perform an expansion of the metric around flat spacetime, $g_{\mu\nu} = \eta_{\mu\nu} + \kappa h_{\mu\nu}$. The expressions for the expansions of the action, the stress tensor and the relevant expectation values are explicitly written down. The computations to first and second order in this expansion are explained in Secs.~\ref{sec:anomaly_k} and~\ref{sec:anomaly_k2}, respectively, and the results for the trace and divergence of the stress tensor displayed.\footnote{We have used the tensor algebra suite xAct~\cite{xact,xpert} to perform the lengthy but straightforward tensor algebra.} We state our conclusions in Sec.~\ref{sec:conclusions}. Some additional computations regarding the one- and two-argument tensorial integrals $\mathcal{I}$ that appear in the computation, the analysis of the necessary counterterms and the expansion of geometrical quantities in terms of the perturbation $h_{\mu\nu}$ are relegated to the appendices.

\subsection*{Conventions}

In the following we will employ the conventions of~\cite{Freedman:2011hp}, in which the metric is $(-,+,+,\cdots)$, the gamma matrices fulfill the usual Clifford algebra, i.e., $\lbrace \gamma^\mu, \gamma^\nu \rbrace = 2 g^{\mu\nu} \1$, and $\bar{\psi} = \mathi \psi^* \gamma^0$. We consistently work in $n$ dimensions in order to employ dimensional regularization, and use the Breitenlohner-Maison scheme~\cite{Breitenlohner:1977hr} for the definition of the chiral matrix $\gamma_*$ in $n$ dimensions. The Riemann tensor is $R^\sigma{}_{\mu\rho\nu} = \partial_\rho \Gamma^\sigma_{\mu\nu} + \cdots$, and the Ricci tensor is obtained as $R_{\mu\nu} = R^\rho{}_{\mu\rho\nu}$. We use geometric units $c = \hbar = 1$ and the totally antisymmetric symbol normalized to $\epsilon_{0123} = 1$. We denote (idempotent) symmetrization of indices by parentheses, e.g., $v^{(a} w^{b)} = \tfrac{1}{2} \left( v^a w^b + v^b w^a \right)$, and antisymmetrization by brackets, e.g., $v^{[a} w^{b]} = \tfrac{1}{2} \left( v^a w^b - v^b w^a \right)$.

\section{Weyl fermions in curved space}
\label{sec:weyl}

First of all, let us introduce left-handed Weyl fermions, which are four-component fermions satisfying
\begin{align}
\label{eq:weyl_fermion_def}
\psi = \mathcal{P}_+ \psi \equiv \frac{1}{2} \left( \1 + \gamma_* \right) \psi, \qquad \bar{\psi} = \bar{\psi} \mathcal{P}_- \equiv \frac{1}{2} \bar{\psi} \left( \1 - \gamma_* \right) \eqend{,}
\end{align}
where $\gamma_*$ is the chiral $\gamma$ matrix satisfying $\gamma_*^2 = \1$, such that the projectors $\mathcal{P}_\pm$ are idempotent: $\mathcal{P}_\pm^2 = \mathcal{P}_\pm$. Since $\bar{\psi} \psi = \bar{\psi} \mathcal{P}_+ \mathcal{P}_- \psi = 0$, no Dirac mass term for Weyl fermions exists. Under a parity transformation, we have $\gamma_* \to - \gamma_*$, and left-handed Weyl fermions are mapped to right-handed ones and vice versa.

To consistently define the chiral matrix $\gamma_*$ in $n$ dimensions, we will use the Breitenlohner-Maison scheme~\cite{Breitenlohner:1977hr} which we are going to explain in more detail in Sec.~\ref{sec:BMprescription}, and which preserves the relation $\gamma_*^2 = \1$ such that eq.~\eqref{eq:weyl_fermion_def} still holds. In a curved $n$-dimensional space, the action for Weyl fermions reads
\begin{equation}
\label{eq:action}
S = - \int \bar{\psi} \gamma^\mu \nabla_\mu \psi \sqrt{-g} \total^n x = - \int \bar{\psi} \mathcal{P}_- \gamma^\mu \nabla_\mu \left( \mathcal{P}_+ \psi \right) \sqrt{-g} \total^n x \eqend{,}
\end{equation}
where in the second expression we have explicitly displayed the projectors. In this expression, the covariant derivative for the fermion includes the spin connection $\omega$,
\begin{equation}
\label{eq:covariant_derivative}
\nabla_\mu \psi \equiv \partial_\mu \psi + \frac{1}{4} \omega_{\mu\rho\sigma} \gamma^{\rho\sigma} \psi \eqend{,}
\end{equation}
and the curved-space $\gamma$ matrices are obtained as usual from the constant flat-space ones by introducing the vielbein, $e_\mu{}^a$, namely as $\gamma^\mu \equiv g^{\mu\nu} e_\nu{}^a \eta_{ab} \gamma^b$. Additionally, we have introduced the higher-order $\gamma$ matrices obtained by antisymmetrization~\cite{Kennedy:1981kp}, $\gamma^{\mu_1 \cdots \mu_k} \equiv \gamma^{[\mu_1} \cdots \gamma^{\mu_k]}$. The spin connection $\omega$ can be expressed in terms of the vielbein~\cite{Freedman:2011hp}:
\begin{equation}
\label{eq:spin_connection}
\omega_{\mu\rho\sigma} = \omega_{\mu[\rho\sigma]} = \eta_{ab} \left( e_\sigma{}^a \partial_{[\mu} e_{\rho]}{}^b - e_\rho{}^a \partial_{[\mu} e_{\sigma]}{}^b + e_\mu{}^a \partial_{[\sigma} e_{\rho]}{}^b \right) \eqend{.}
\end{equation}

From the action \eqref{eq:action}, one can compute the associated symmetric stress tensor~\cite{Forger:2003ut} (with the projectors explicitly displayed)
\begin{equation}
\label{eq:EMtensor}
T^{\mu\nu} = \frac{1}{2} \bar{\psi} \mathcal{P}_- \gamma^{(\mu} \overleftrightarrow{\nabla}^{\nu)} \mathcal{P}_+ \psi + \frac{1}{2} g^{\mu\nu} \bar{\psi} \mathcal{P}_- \gamma^\rho \overleftrightarrow{\nabla}_\rho \mathcal{P}_+ \psi \eqend{,}
\end{equation}
where we have introduced $\overleftrightarrow{\nabla}_\mu \equiv \nabla_\mu - \overleftarrow{\nabla}_\mu$, and $\overleftarrow{\nabla}_\mu$ acts according to $\bar{\psi} \overleftarrow{\nabla}_\mu  \equiv \partial_\mu \bar{\psi} - \frac{1}{4} \bar\psi\, \omega_{\mu\rho\sigma} \gamma^{\rho\sigma}$. As is well known (and can be checked easily using that $\left[ \nabla_\mu, \nabla_\nu \right] \psi = \tfrac{1}{4} R_{\mu\nu\rho\sigma} \gamma^{\rho\sigma} \psi$~\cite{Freedman:2011hp}), classically this tensor is conserved and its trace vanishes on-shell, since the action for a Weyl fermion is conformally invariant. Furthermore, the second term in the RHS of \eqref{eq:EMtensor} vanishes on-shell. As we will see the situation can change at the quantum level, giving rise to the trace anomaly.

\subsection{The Breitenlohner-Maison prescription}
\label{sec:BMprescription}

The main problem when dealing with dimensional regularization of chiral theories is related to the appropriate extension of $\gamma_*$ from $n = 4$ to general $n$ dimensions. This comes from its definition in terms of the totally antisymmetric symbol $\epsilon$, which is intrinsically a four-dimensional object; in formulas, $\gamma_*^{n=4} = - \frac{\mathi}{4!}\epsilon_{\mu\nu\rho\sigma} \gamma^\mu \gamma^\nu \gamma^\rho \gamma^\sigma$. Indeed, a contradiction can be explicitly seen to emerge from the following two properties, valid in $n = 4$ dimensions:
\begin{equations}
\tr\left( \gamma_* \gamma_\mu \gamma_\nu \gamma_\rho \gamma_\sigma \right) &= - \mathi \, \epsilon_{\mu\nu\rho\sigma} \tr \1 \eqend{,} \label{eq:g5fourg} \\
\lbrace \gamma_\mu, \gamma_* \rbrace &= 0 \eqend{.} \label{eq:g5ganticommute}
\end{equations}
As is well-known, if we take eq.~\eqref{eq:g5ganticommute} as granted in general $n$ dimensions, together with the cyclicity of the trace and $\gamma_*^2 = \1$, it follows that
\begin{equation}
n \tr \gamma_* = \tr\left( \gamma_* \gamma^\alpha \gamma_\alpha \right) = - \tr\left( \gamma^\alpha \gamma_* \gamma_\alpha \right) = - \tr\left( \gamma_* \gamma_\alpha \gamma^\alpha \right) = - n \tr \gamma_*
\end{equation}
and thus $n \tr \gamma_* = 0$, then
\begin{splitequation}
n \tr\left( \gamma_* \gamma_\mu \gamma_\nu \right) &= \tr\left( \gamma_* \gamma^\alpha \gamma_\alpha \gamma_\mu \gamma_\nu \right) = - \tr\left( \gamma_* \gamma_\alpha \gamma_\mu \gamma_\nu \gamma^\alpha \right) \\
&= \ldots = - 2 \tr\left( \gamma_* \gamma_\nu \gamma_\mu \right) + 2 \tr\left( \gamma_* \gamma_\mu \gamma_\nu \right) - n \tr\left( \gamma_* \gamma_\mu \gamma_\nu \right) \\
&= - 4 g_{\mu\nu} \tr\gamma_* - (n-4) \tr\left( \gamma_* \gamma_\mu \gamma_\nu \right)
\end{splitequation}
and thus $n (n-2) \tr\left( \gamma_* \gamma_\mu \gamma_\nu \right) = 0$, and analogously
\begin{equation}
n (n-2) (n-4) \tr\left( \gamma_* \gamma_\mu \gamma_\nu \gamma_\rho \gamma_\sigma \right) = 0 \eqend{.}
\end{equation}
which contradicts formula~\eqref{eq:g5fourg} for $n \neq 4$.

The proposal of Breitenlohner and Maison~\cite{Breitenlohner:1977hr} to solve this problem, based upon~\cite{tHooft:1972tcz}, consists in splitting the $n$-dimensional Minkowski space in the product of a four-dimensional and an $(n-4)$-dimensional one, and keeping the anticommutativity property~\eqref{eq:g5ganticommute} only for the four-dimensional subspace. That is, we consider in $n$ dimensions the usual metric $\eta_{\mu\nu}$, gamma matrices $\gamma_\mu$ and momenta $p_\mu$, and decompose them into a four-dimensional part (denoted with an overbar) and an $(n-4)$-dimensional part (denoted with a hat):\footnote{In general, one has to perform this decomposition for the indices, such that also mixed quantities can appear. However, the Minkowski metric is diagonal in $n$ dimensions, and no mixed metric exists.}
\begin{equation}
\eta_{\mu\nu} = \bar{\eta}_{\mu\nu} + \hat{\eta}_{\mu\nu} \eqend{,} \quad \gamma_\mu = \bar{\gamma}_\mu + \hat{\gamma}_\mu \eqend{,} \quad p_\mu = \bar{p}_\mu + \hat{p}_\mu \eqend{,} \quad \ldots
\end{equation}
While all objects are defined in the full $n$-dimensional Minkowski space, they vanish whenever their indices do not pertain to the corresponding subspace, such that we have for example the following relations:
\begin{splitequation}
\eta_\mu{}^\nu \hat{\eta}_{\nu\rho} &= \hat{\eta}_{\mu\nu} \hat{\eta}^\nu{}_\rho = \hat{\eta}_{\mu\rho} \eqend{,} \quad \hat\eta_{\mu\nu} = \hat\eta_{\nu\mu} \eqend{,} \\
\bar{\eta}^{\mu\nu} \hat{\eta}_{\nu\rho} &= 0 \eqend{,} \quad \bar{\eta}_\mu{}^\nu p_\nu = \bar{p}_\mu \eqend{,} \quad \hat{\eta}_\mu{}^\nu \gamma_\nu = \hat{\gamma}_\mu \eqend{,}
\end{splitequation}
and many more. The totally antisymmetric symbol $\epsilon_{\mu\nu\rho\sigma}$ is a purely four-dimensional object: $\epsilon_{\mu\nu\rho\sigma} = \bar{\epsilon}_{\mu\nu\rho\sigma}$ and $\hat{\eta}^{\alpha\mu} \epsilon_{\mu\nu\rho\sigma} = 0$. The definition of the chiral matrix $\gamma_*$ is then the same as in four dimensions, namely
\begin{equation}
\label{eq:gamma_star_def}
\gamma_* \equiv - \frac{\mathi}{4!} \epsilon_{\mu\nu\rho\sigma} \gamma^\mu \gamma^\nu \gamma^\rho \gamma^\sigma = - \frac{\mathi}{4!} \epsilon_{\mu\nu\rho\sigma} \gamma^{[\mu\nu\rho\sigma]} \eqend{.}
\end{equation}
From this, one can easily derive the following identities~\cite{Breitenlohner:1977hr}:
\begin{equations}
\lbrace \gamma_\mu, \hat\gamma_\nu \rbrace &= \lbrace \hat\gamma_\mu, \hat\gamma_\nu \rbrace = 2 \hat \eta_{\mu\nu} \1 \eqend{,} \\
\hat\eta_\mu{}^\mu &= n-4 \eqend{,} \label{eq:hateta_trace} \\
\lbrace \gamma_\mu, \gamma_* \rbrace &= \lbrace \hat\gamma_\mu, \gamma_* \rbrace = 2 \hat\gamma_\mu \gamma_* \eqend{,} \\
\left[ \gamma_\mu, \gamma_* \right] &= \left[ \bar{\gamma}_\mu, \gamma_* \right] = 2 \bar\gamma_\mu \gamma_* \eqend{,} \\
\gamma_*^2 &= \1 \eqend{,} \\
\mathcal{P}_\pm \gamma^\mu \mathcal{P}_\mp &= \mathcal{P}_\pm \bar\gamma^\mu \eqend{,} \label{eq:gamma_sandwich_projector}
\end{equations}
where the last formula is a consequence of the definition of the projectors.

Since this definition breaks $n$-dimensional Lorentz covariance for chiral objects, spurious non-covariant terms may appear in the calculation, which can however be canceled with (finite) non-covariant counterterms~\cite{Bonneau:1980ya,Martin:1999cc,SanchezRuiz:2002xc,Belusca-Maito:2020ala}.

\subsection{Perturbative expansion around Minkowski space}
\label{sec:expansion}

In order to study the trace anomaly, we consider fluctuations $h_{\mu\nu}$ of the metric around Minkowski space:
\begin{align}
g_{\mu\nu} = \eta_{\mu\nu} + \kappa h_{\mu\nu} \eqend{,}
\end{align}
with $\kappa$ a parameter that helps to keep track of the expansion order. The vielbein, the inverse metric, etc. of course change accordingly. Choosing symmetric gauge for the vielbein~\cite{Woodard:1984sj}, up to and including quadratic contributions in $h$ they explicitly read
\begin{equations}
\label{eq:geometric_expansion}
e^\mu{}_a &= e^{(0)}{}^\rho{}_a \left( \eta^\mu{}_\rho - \frac{1}{2} \kappa h^\mu{}_\rho + \frac{3}{8} \kappa^2 h^{\mu\sigma} h_{\sigma\rho} \right) + \bigo{\kappa^3} \eqend{,} \\
e_\mu{}^a &= e^{(0)}{}^a{}_\rho \left( \eta^{\mu\rho} + \frac{1}{2} \kappa h^{\mu\rho} - \frac{1}{8} \kappa^2 h^{\mu\sigma} h_\sigma{}^\rho \right) + \bigo{\kappa^3} \eqend{,} \\
g^{\mu\nu} &= \eta^{\mu\nu} - \kappa h^{\mu\nu} + \kappa^2 h^{\mu\rho} h_\rho{}^\nu + \bigo{\kappa^2} \eqend{,} \\
\sqrt{-g} &= 1 + \frac{\kappa}{2} h^\rho{}_\rho + \frac{\kappa}{8} \left( h^\rho{}_\rho h^\sigma{}_\sigma - 2 h^{\rho\sigma} h_{\rho\sigma} \right) + \bigo{\kappa^3} \eqend{,}
\end{equations}
where $e^{(0)}{}^\rho{}_a = \delta^\rho_a$ is the vielbein in the background Minkowski space. It is important to notice that in the right-hand side of \eqref{eq:geometric_expansion} we have used the flat metric $\eta$ to raise indices, even if we are writing them with the same Greek indices as if we had used the full metric. This is the convention that will be employed in the following. For notational convenience, we will also introduce a notation for the expansions in $h$: Given a quantity $A$, we set
\begin{align}
A &\equiv A_{(0)} + \kappa A_{(1)} + \kappa^2 A_{(2)} + \bigo{\kappa^3} \eqend{.}
\end{align}
In particular, we will expand the vacuum expectation value (VEV) of the stress tensor in the curved spacetime with metric $g$ as
\begin{align}
\label{eq:expansion_VEV_EM}
\expect{ T^{\mu\nu}(x) }_g &= \expect{ T^{\mu\nu}(x) }_{(0)} + \kappa \expect{ T^{\mu\nu}(x) }_{(1)} + \kappa^2 \expect{ T^{\mu\nu}(x) }_{(2)} + \bigo{\kappa^3} \eqend{,}
\end{align}
where the VEV's on the right-hand side are all computed in the flat Minkowski background (which will always be the case in the following, except if otherwise indicated by a subindex $g$ like on the left-hand side).

Expanding also the action $S$ in a free part $S_{(0)}$ and interactions $S_{(1)}$, $S_{(2)}$, ... between the fermions and the perturbation $h_{\mu\nu}$, the Gell-Mann--Low formula gives
\begin{equation}
\expect{ T^{\mu\nu}(x) }_g = \frac{\expect{ T^{\mu\nu}(x) \exp\left[ \mathi \left( \kappa S_{(1)} + \kappa^2 S_{(2)} \right) \right]}}{\expect{ \exp\left[ \mathi \left( \kappa S_{(1)} + \kappa^2 S_{(2)} \right) \right]}} + \bigo{\kappa^3} \eqend{,}
\end{equation}
where the expectation values without suffix on the right-hand side denote expectation values in the free theory in Minkowski spacetime with action $S_{(0)}$, see eqs.~\eqref{eq:actionexpansion_s0} and~\eqref{eq:actions0_new}. Expanding in $\kappa$ and comparing with the expansion~\eqref{eq:expansion_VEV_EM}, we obtain
\begin{equations}[eq:coefficients_VEV]
\expect{ T^{\mu\nu}(x) }_{(1)} &= \mathi \expect{ S_{(1)} T^{\mu\nu}_{(0)}(x) } \eqend{,} \\
\expect{ T^{\mu\nu}(x) }_{(2)} &= - \frac{1}{2} \expect{ S_{(1)}^2 T^{\mu\nu}_{(0)}(x) } + \mathi \expect{ S_{(2)} T^{\mu\nu}_{(0)}(x) } + \mathi \expect{ S_{(1)} T^{\mu\nu}_{(1)}(x) } \eqend{.}
\end{equations}
We have already simplified this expression by noting that in dimensional regularization, the VEV of a quantity which depends on just one position variable vanishes since all fields are massless: $\expect{ T^{\mu\nu}_{(k)}(x) } = 0 = \expect{ S_{(k)}(x) }$ for $k = 0,1,2$. Before giving the explicit form of the coefficients, we further compactify the notation and define
\begin{equations}[eq:structures]
\Psi^{\mu\nu} &\equiv \bar\psi \mathcal{P}_- \gamma^\mu \mathcal{P}_+ \partial^\nu \psi - \partial^\nu \bar\psi \mathcal{P}_- \gamma^\mu \mathcal{P}_+ \psi \eqend{,} \\
J^{\mu\nu\rho} &\equiv \bar\psi \mathcal{P}_- \gamma^{\mu\nu\rho} \mathcal{P}_+ \psi \eqend{,}
\end{equations}
as well as the traces $\Psi \equiv \Psi^\mu{}_\mu$ and $h \equiv h^\mu{}_\mu$. In this way, we obtain the coefficients in the expansion of the stress tensor,
\begin{equations}[eq:emtensorexpansion]
T^{\mu\nu}_{(0)} &= \frac{1}{2} \left[ \Psi^{(\mu\nu)} - \Psi \eta^{\mu\nu} \right] \eqend{,} \label{eq:emtensorexpansion_0} \\
T^{\mu\nu}_{(1)} &= \frac{1}{4} \left[ 2 \Psi h^{\mu\nu} + h_{\alpha\beta} \Psi^{\alpha\beta} \eta^{\mu\nu} - 2 h^{\alpha(\mu} \Psi^{\nu)}{}_\alpha - h^{\alpha(\mu} \Psi_\alpha{}^{\nu)} - J^{\alpha\beta(\mu} \partial_\alpha h^{\nu)}{}_\beta \right] \eqend{,} \label{eq:emtensorexpansion_1} \\
\begin{split}
T^{\mu\nu}_{(2)} &= \frac{1}{16} \Big[ 4 h^{\rho(\mu} h^{\nu)\sigma} - 4 h_{\rho\sigma} h^{\mu\nu} + 8 h^{\alpha(\mu} \left( \delta^{\nu)}_\rho h_{\alpha\sigma} - h^{\nu)}{}_\alpha \eta_{\rho\sigma} \right) \\
&\qquad+ 3 h_{\alpha\rho} \left( h^{\alpha(\mu} \delta^{\nu)}_\sigma - h_\sigma{}^\alpha \eta^{\mu\nu} \right) \Big] \Psi^{\rho\sigma} + \frac{1}{16} \Big[ \left( 4 h^{\alpha(\mu} \eta^{\nu)\rho} - \eta^{\mu\nu} h^{\alpha\rho} \right) \partial^\sigma h_\alpha{}^\tau \\
&\hspace{3em}+ 2 h^{\mu\rho} \partial^\sigma h^{\nu\tau} - h^{\alpha\sigma} \eta^{\rho(\mu} \left( \partial^{\nu)} h_\alpha{}^\tau - 2 \partial_\alpha h^{\nu)}{}_\tau + 2 \partial_\tau h^{\nu)}{}_\alpha \right) \Big] J_{\rho\sigma\tau} \eqend{,} \label{eq:emtensorexpansion_2}
\end{split}
\end{equations}
and those corresponding to the action,
\begin{equations}[eq:actionexpansion]
S_{(0)} &= - \frac{1}{2} \int \Psi \total^n x \eqend{,} \label{eq:actionexpansion_s0} \\
S_{(1)} &= \frac{1}{4} \int \left( h_{\alpha\beta} \Psi^{\alpha\beta} - h \Psi \right) \total^n x \eqend{,} \label{eq:actionexpansion_s1} \\
S_{(2)} &= \frac{1}{16} \int \left[ \left( 2 h h_{\alpha\beta} - 3 h_\alpha{}^\delta h_{\beta\delta} \right) \Psi^{\alpha\beta} + \left( 2 h_{\alpha\beta} h^{\alpha\beta} - h^2 \right) \Psi + h_\alpha{}^\delta \partial_\gamma h_{\beta\delta} J^{\alpha\beta\gamma} \right] \total^n x \eqend{.} \label{eq:actionexpansion_s2}
\end{equations}

Note that while in $n = 4$ dimensions, we can dispense with some of the projectors $\mathcal{P}_\pm$ since $\gamma_*$ anticommutes with all other $\gamma$ matrices, this is not possible in $n$ dimensions. However, the differential operator appearing in the free action $S_{(0)}$~\eqref{eq:actionexpansion_s0} is clearly not invertible because of the projectors. To remedy this problem and determine the free propagator, we add a free right-handed Weyl fermion $\mathcal{P}_- \psi$ to the action~\cite{AlvarezGaume:1983cs,GamboaSaravi:1988wf}, which is also the route followed by Bonora et al.~\cite{Bonora:2014qla,Bonora:2015nqa}. Since the stress tensor~\eqref{eq:emtensorexpansion} does not depend on this right-handed fermion [which is ensured by the projectors in eq.~\eqref{eq:structures}], and no interaction terms are added for it, this does not change the expectation value of the stress tensor. Instead of~\eqref{eq:actionexpansion_s0}, we thus have the free action
\begin{equation}
\label{eq:actions0_new}
S_{(0)}(x) = - \int \bar\psi \gamma^\mu \partial_\mu \psi \total^n x \eqend{,}
\end{equation}
which results in the fermion propagator
\begin{equation}
\label{eq:fermion_prop}
G_0(x,y) = - \mathi \expect{ \psi(x) \bar{\psi}(y) } = \int \frac{\gamma^\rho k_\rho}{k^2 - \mathi 0} \mathe^{\mathi k (x-y)} \dn k \eqend{.}
\end{equation}
If desired, from the expansion of the action~\eqref{eq:actionexpansion} one can derive all Feynman rules for an expansion in Feynman diagrams. However, we will directly use the expansion~\eqref{eq:coefficients_VEV} together with Wick's theorem to evaluate the free-theory VEV's.

The information about chirality is stored thus in the interaction terms, in which every fermion possesses its own projector. This is in contrast with~\cite{Bonora:2014qla,Bonora:2015nqa}, where only one projector appears in the three- or four-point interaction vertices. Both interactions are equal in $n = 4$ dimensions, but not in $n$ dimensions using the Breitenlohner--Maison scheme, and for a consistent implementation of dimensional regularization one needs to work from the beginning in $n$ dimensions. This can be seen, for example, in the computation of loop corrections in inflation, where the contribution from logarithms of the scale factor $a$ is crucial to ensure de Sitter invariance of the result~\cite{Weinberg:2006ac,Adshead:2008gk,Senatore:2009cf,Weinberg:2010wq,Frob:2012ui}. These logarithms are obtained from the determinant of the de Sitter metric $a^n \approx a^4 \left[ 1 + (n-4) \ln a \right]$ and give a finite contribution when multiplying the pole in $n-4$, but are missed if one starts in $n = 4$ dimensions and only dimensionally regularizes the momentum integrations.

Lastly, consider the expressions for the trace $\mathcal{T} \equiv \expect{ T^\mu{}_\mu }_g$ and the divergence $\mathcal{D}^\nu \equiv \expect{ \nabla_\mu T^{\mu\nu} }_g$ of the VEV of the stress tensor in the curved spacetime with metric $g$. Up to second order, their expansion coefficients are given by
\begin{equations}
\mathcal{T}_{(1)}(x) &= \eta_{\mu\nu} \expect{ T^{\mu\nu}(x) }_{(1)} \eqend{,} \label{eq:tr_orderh} \\
\mathcal{T}_{(2)}(x) &= \eta_{\mu\nu} \expect{ T^{\mu\nu}(x) }_{(2)} + h_{\mu\nu} \expect{ T^{\mu\nu}(x) }_{(1)} \eqend{,} \label{eq:tr_orderh2} \\
\mathcal{D}^\nu_{(1)}(x) &= \partial_\mu \expect{ T^{\mu\nu}(x) }_{(1)} \eqend{,} \label{eq:div_orderh} \\
\mathcal{D}^\nu_{(2)}(x) &= \partial_\mu \expect{ T^{\mu\nu}(x) }_{(2)} + \frac{1}{2} \left( 2 \partial_\rho h^\nu{}_\mu - \partial^\nu h_{\rho\mu} + \delta^\nu_\mu \partial_\rho h \right) \expect{ T^{\mu\rho}(x) }_{(1)} \eqend{.} \label{eq:div_orderh2}
\end{equations}

\section{The trace anomaly at \texorpdfstring{$\bigo{\kappa}$}{O(\textkappa)}}
\label{sec:anomaly_k}

The first order computations are straightforward and will serve as a warm-up for the second order ones. A direct consequence of the results in the previous section is that the only relevant quantity at this order is the VEV
\begin{splitequation}
\label{eq:VEV_2pf}
&\expect{ \Psi^{\mu\nu}(x) \Psi^{\rho\sigma}(y) } = \tr\left( \mathcal{P}_+ \gamma^\tau \mathcal{P}_- \gamma^\mu \mathcal{P}_+ \gamma^\lambda \mathcal{P}_- \gamma^\rho \right) \\
&\hspace{4em}\times \iint (2q+p)^\nu (2q+p)^\sigma \frac{(p+q)_\tau}{(p+q)^2-\mathi 0} \frac{q_\lambda}{q^2-\mathi 0} \mathe^{\mathi p(x-y)} \dn p \dn q \eqend{.}
\end{splitequation}
We will call this the two-point function\footnote{This name arises from the fact that $\Psi^{\mu\nu}$, upon symmetrization and subtraction of a trace, corresponds to the flat $T^{\mu\nu}_{(0)}$, which is sometimes used as an element in the expansion.}, and the right-hand side has been written in a way such that one can recognize the emergence of the tensorial integrals
\begin{align}
\label{eq:Ip}
\mathcal{I}^{\mu_1 \cdots \mu_m}(p) \equiv \int \frac{q^{\mu_1} \cdots q^{\mu_m}}{(q^2-\mathi 0) [ (q+p)^2 - \mathi 0 ]} \dn q \eqend{.}
\end{align}
Using the Passarino--Veltman procedure~\cite{Passarino:1978jh}, these integrals can be reduced to a product of the scalar integral $\mathcal{I}(p)$ times a tensorial factor depending on the external momentum $p$ and the flat metric, and we collect the formulas relevant to the present computation in App.~\ref{app:integrals_1}. After a straightforward computation, we obtain the expression for the two-point function
\begin{splitequation}
\label{eq:VEV_2pf_integrated}
&\expect{ \Psi^{\mu\nu}(x) \Psi^{\rho\sigma}(y) } = \tr\left( \mathcal{P}_+ \gamma_\tau \mathcal{P}_- \gamma^\mu \mathcal{P}_+ \gamma_\lambda \mathcal{P}_- \gamma^\rho \right) \\
&\hspace{4em}\times \int \frac{\mathcal{I}(p)}{4 (n^2-1)} \left[ - (n+1) p^\lambda p^\tau \Pi^{\nu\sigma}(p) + 3 \Pi^{(\lambda\tau}(p) \Pi^{\nu\sigma)}(p) \right] \mathe^{\mathi p(x-y)} \dn p
\end{splitequation}
in terms of the projector onto the space orthogonal to the vector $p$:
\begin{align}
\Pi^{\mu\nu}(p) \equiv p^\mu p^\nu - \eta^{\mu\nu} p^2 \eqend{.}
\end{align}

\subsection{The original Breitenlohner--Maison scheme}

The expression~\eqref{eq:VEV_2pf_integrated} can be further simplified by computing the $\gamma$ matrix trace
\begin{align}
\label{eq:1_spinorial}
\tr\left( \mathcal{P}_+ \gamma^\tau \mathcal{P}_- \gamma^\mu \mathcal{P}_+ \gamma^\lambda \mathcal{P}_- \gamma^\rho \right) = 2 \left( \bar\eta^{\tau\mu} \bar\eta^{\lambda\rho} - \bar\eta^{\tau\lambda} \bar\eta^{\mu\rho} + \bar\eta^{\tau\rho} \bar\eta^{\mu\lambda} \right) - 2 \mathi \epsilon^{\tau\mu\lambda\rho} \eqend{,}
\end{align}
where we used eq.~\eqref{eq:gamma_sandwich_projector}. According to the Breitenlohner--Maison scheme~\cite{Breitenlohner:1977hr,Bonneau:1980ya,Rufa:1990hg}, external momenta such as $p^\mu$ (i.e., the ones not involved in the loop integration) are always taken in the four-dimensional spacetime. Performing the contractions we obtain for the two-point function
\begin{splitequation}
\label{eq:VEV_2pf_integrated_b}
\expect{ \Psi^{\mu\nu}(x) \Psi^{\rho\sigma}(y) } &= \frac{1}{n^2-1} \int \mathcal{I}(p) \Big[ \bar{\Pi}^{\mu\nu}(p) \bar{\Pi}^{\rho\sigma}(p) + \bar{\Pi}^{\mu\sigma}(p) \bar{\Pi}^{\nu\rho}(p) \\
&\hspace{-2em}+ \frac{n+2}{2} p^2 \bar\eta^{\mu\rho} \Pi^{\nu\sigma}(p) + p^2 \bar\eta^{\mu\rho} \bar\Pi^{\nu\sigma}(p) - n p^\rho p^\mu \Pi^{\nu\sigma}(p) \Big] \mathe^{\mathi p (x-y)} \dn[4] p
\end{splitequation}
with
\begin{equation}
\label{eq:barpi_def}
\bar\Pi^{\mu\nu} \equiv p^\mu p^\nu - \bar\eta^{\mu\nu} p^2 = \Pi^{\mu\nu} + \hat\eta^{\mu\nu} p^2 \eqend{.}
\end{equation}
For the ($n$-dimensional) trace, we obtain
\begin{equations}[eq:2pf_traces]
\expect{ \Psi^{\mu\nu}(x) \Psi(y) } &= \eta_{\rho\sigma} \expect{ \Psi^{\mu\nu}(x) \Psi^{\rho\sigma}(y) } = \frac{n-4}{2 (n^2-1)} \int \mathcal{I}(p) \, p^2 \bar{\Pi}^{\mu\nu}(p) \mathe^{\mathi p (x-y)} \dn[4] p \eqend{,} \raisetag{1em} \\
\expect{ \Psi(x) \Psi^{\rho\sigma}(y) } &= \frac{n-4}{2 (n^2-1)} \int \mathcal{I}(p) \, p^2 \bar{\Pi}^{\rho\sigma}(p) \mathe^{\mathi p (x-y)} \dn[4] p \eqend{,} \\
\expect{ \Psi(x) \Psi(y) } &= - \frac{3 (n-4)}{2 (n^2-1)} \int \mathcal{I}(p) \, (p^2)^2 \mathe^{\mathi p (x-y)} \dn[4] p \eqend{,}
\end{equations}
and taking all together the regularized expression for the first-order contribution to the VEV of the stress tensor follows:
\begin{splitequation}
\label{eq:VEV1_reg}
\expect[reg]{ T^{\mu\nu}(x) }_{(1)} &= \frac{\mathi}{8 (n^2-1)} \iint \mathcal{I}(p) \Big[ \bar{\Pi}^{\mu\nu}(p) \bar{\Pi}^{\rho\sigma}(p) - (n-1) \bar{\Pi}^{\rho(\mu}(p) \bar{\Pi}^{\nu)\sigma}(p) \\
&\hspace{3em}- \frac{n-4}{2} p^2 \left( \bar\eta^{\rho(\mu} \bar{\Pi}^{\nu)\sigma}(p) + \bar{\Pi}^{\mu\nu}(p) \bar{\eta}^{\rho\sigma} + \eta^{\mu\nu} \bar{\Pi}^{\rho\sigma}(p) + 3 \eta^{\mu\nu} \bar{\eta}^{\rho\sigma} p^2 \right) \Big] \\
&\hspace{4em}\times \mathe^{\mathi p (x-y)} \dn[4] p h_{\rho\sigma}(y) \total^4 y \eqend{,} \raisetag{2em}
\end{splitequation}
where also the external field $h_{\rho\sigma}$ must be taken purely four-dimensional~\cite{Breitenlohner:1977hr,Bonneau:1980ya,Rufa:1990hg}. Due to the breaking of $n$-dimensional Lorentz covariance in the Breitenlohner--Maison scheme, this regularized stress tensor is now neither conserved nor traceless. Instead, we obtain
\begin{splitequation}
\label{eq:VEV1_reg_trace}
\mathcal{T}_{(1)}^\text{reg}(x) &= \eta_{\mu\nu} \expect[reg]{ T^{\mu\nu}(x) }_{(1)} \\
&= - \frac{\mathi (n-4)}{16 (n+1)} \iint \mathcal{I}(p) \, p^2 \left[ \bar{\Pi}^{\rho\sigma}(p) + 3 \bar{\eta}^{\rho\sigma} p^2 \right] \mathe^{\mathi p (x-y)} \dn[4] p h_{\rho\sigma}(y) \total^4 y
\end{splitequation}
and
\begin{splitequation}
\label{eq:VEV1_reg_div}
\mathcal{D}^{\nu}_{\text{reg},\,(1)}(x) &= \partial_\mu \expect[reg]{ T^{\mu\nu}(x) }_{(1)} = \frac{n-4}{32 (n^2-1)} \iint \mathcal{I}(p) \, p^2 \\
&\hspace{4em}\times \left[ p^\rho \bar{\Pi}^{\nu\sigma}(p) + 2 p^\nu \bar{\Pi}^{\rho\sigma}(p) + 6 p^\nu \bar{\eta}^{\rho\sigma} p^2 \right] \mathe^{\mathi p (x-y)} \dn[4] p h_{\rho\sigma}(y) \total^4 y \eqend{,}
\end{splitequation}
where an overall coefficient $(n-4)$ has appeared, indicating that this can be fixed by a finite renormalization. Of course, we first have to renormalize the VEV of the stress tensor itself, for which we need the explicit expression of the scalar integral $\mathcal{I}(p)$. This is a standard computation with result~\cite{smirnov,Godazgar:2018boc}
\begin{align}
\label{eq:scalar_integral}
\mathcal{I}(p) = \frac{\mathi}{16 \pi^2} \left[ \diver + 2 - \ln\left( \frac{p^2 - \mathi 0}{\mu^2} \right) \right] + \bigo{n-4} \eqend{,}
\end{align}
where $\mu$ is a reference scale and
\begin{align}
\diver \equiv - \frac{2}{n-4} - \gamma - \log\left( \frac{\mu^2}{4\pi} \right) \eqend{.}
\end{align}
We renormalize in the \MS scheme, which is defined by subtracting $\diver$ only and setting all explicit factors of $n$ in the integrand~\eqref{eq:VEV1_reg} to $4$. This results in
\begin{splitequation}
\label{eq:VEV1_div}
\expect[div]{ T^{\mu\nu}(x) }_{(1)} &= - \frac{\diver}{120\cdot 16 \pi^2} \iint \Big[ \bar{\Pi}^{\mu\nu}(p) \bar{\Pi}^{\rho\sigma}(p) - 3 \bar{\Pi}^{\rho(\mu}(p) \bar{\Pi}^{\nu)\sigma}(p) \Big] \\
&\hspace{4em}\times \mathe^{\mathi p (x-y)} \dn[4] p h_{\rho\sigma}(y) \total^4 y \\
&= - \frac{\diver}{120\cdot 16 \pi^2} \left( 6 \bar\nabla^2 \bar{R}^{\mu\nu} - 2 \bar\nabla^\mu \bar\nabla^\nu \bar{R} - \bar{g}^{\mu\nu} \bar\nabla^2 \bar{R} \right)_{(1)}(x) \eqend{,}
\end{splitequation}
where to arrive at the second expression we traded the $p$ variables for derivatives acting on $h$ and compared the resulting expression with the expansions of geometrical quantities given in App.~\ref{app:geometric_expansion}. We leave to App.~\ref{app:counterterm} the proof that this divergent part of the stress tensor VEV (together with the second-order result) corresponds to a counterterm that can be added to the action; the renormalized stress tensor VEV is then given by
\begin{splitequation}
\label{eq:VEV1_ren}
\expect[ren]{ T^{\mu\nu}(x) }_{(1)} &= \expect[reg]{ T^{\mu\nu}(x) }_{(1)} - \expect[div]{ T^{\mu\nu}(x) }_{(1)} \\
&= - \frac{1}{120 \cdot 16 \pi^2} \iint \left[ \frac{46}{15} - \ln\left( \frac{p^2 - \mathi 0}{\mu^2} \right) \right] \\
&\hspace{4em}\times \left[ \bar{\Pi}^{\mu\nu}(p) \bar{\Pi}^{\rho\sigma}(p) - 3 \bar{\Pi}^{\mu(\rho}(p) \bar{\Pi}^{\sigma)\nu}(p) \right] \mathe^{\mathi p (x-y)} \dn[4] p h_{\rho\sigma}(y) \total^4 y \\
&\hspace{0.95em}- \frac{1}{120 \cdot 16 \pi^2} \iint \Big[ p^2 \left[ \bar\eta^{\rho(\mu} \bar\Pi^{\nu)\sigma}(p) + \bar{\Pi}^{\mu\nu}(p) \bar\eta^{\rho\sigma} + \bar\eta^{\mu\nu} \bar{\Pi}^{\rho\sigma}(p) + 3 \bar\eta^{\mu\nu} \bar\eta^{\rho\sigma} p^2 \right] \\
&\hspace{5em}+ 2 \bar{\Pi}^{\rho(\mu}(p) \bar{\Pi}^{\nu)\sigma}(p) \Big] \mathe^{\mathi p (x-y)} \dn[4] p h_{\rho\sigma}(y) \total^4 y \eqend{,} \raisetag{2em}
\end{splitequation}
where everything is now in four dimensions. While the terms on the first line are both conserved and traceless, the terms on the second line are not. However, since they are local, they can be removed by a further finite renormalization. Namely, for the terms in brackets (without the prefactor) we obtain
\begin{splitequation}
\label{eq:VEV1_finite_ct}
&2 \partial^\mu \partial^\nu \partial^\rho \partial^\sigma h_{\rho\sigma}(x) + \partial^\mu \partial^\nu \partial^2 h(x) - 3 \partial^2 \partial^{(\mu} \partial_\rho h^{\nu)\rho}(x) \\
&\qquad+ \partial^2 \left[ \bar\eta^{\mu\nu} \partial^\rho \partial^\sigma h_{\rho\sigma}(x) + \partial^2 h^{\mu\nu}(x) + \bar\eta^{\mu\nu} \partial^2 h(x) \right] \\
&= \frac{1}{2} \frac{\delta}{\delta h_{\mu\nu}(x)} \int \bigg[ \partial^2 h^{\rho\sigma} \partial^2 h_{\rho\sigma} + \left( \partial^2 h \right)^2 + 3 \partial_\rho h^{\rho\alpha} \partial^2 \partial^\sigma h_{\sigma\alpha} \\
&\hspace{8em}+ 2 \partial^2 h \partial_\rho \partial_\sigma h^{\rho\sigma} + 2 \left( \partial_\rho \partial_\sigma h^{\rho\sigma} \right)^2 \bigg] \total^n y \eqend{,}
\end{splitequation}
such that adding this term (with the correct prefactor) to the action, we can completely cancel the terms on the second line of the first-order VEV~\eqref{eq:VEV1_ren}, obtaining a conserved and traceless stress tensor at first order:
\begin{equation}
\mathcal{T}^\text{ren}_{(1)}(x) = 0 = \mathcal{D}^\nu_{\text{ren},\, (1)}(x) \eqend{.}
\end{equation}
Nevertheless, the finite counterterms~\eqref{eq:VEV1_finite_ct} are not covariant, as can be easily checked by comparing them with the expansions of geometrical quantities given in App.~\ref{app:geometric_expansion}.

\subsection{The modified Breitenlohner--Maison scheme}

Since in the Breitenlohner--Maison scheme, the quantum action principle holds~\cite{Breitenlohner:1977hr,Bonneau:1980ya,Rufa:1990hg}, it follows from the results of algebraic renormalization (both in flat~\cite{Piguet:1995er,Sibold:2000mp} and curved~\cite{Hollands:2007zg,Fredenhagen:2011mq,Frob:2018buw} spacetimes) that it is always possible to add finite counterterms to the action to cancel spurious non-covariant terms that arise from the breaking of $n$-dimensional Lorentz covariance. In this way, the same criterion for the appearance of anomalies holds as in any other consistent scheme, namely that anomalies are non-trivial solutions of the Wess--Zumino consistency conditions~\cite{Wess:1971yu}, or more general, elements of a certain cohomology class of the BRST differential~\cite{Bonora:1983ff,Piguet:1995er,Sibold:2000mp}, which for diffeomorphisms is empty in four dimensions~\cite{Barnich:1995ap} (except for possible topological terms that are relevant here).

Therefore, and as suggested in~\cite{Jegerlehner:2000dz}, it is convenient to modify the Breitenlohner--Maison scheme in the parity-even sector since the axial anomaly (as well as a possible Pontryagin density in the trace anomaly) are odd under parity transformations. Namely, for traces containing an even number of the chiral matrix $\gamma_*$, one may use the naive anticommutation rules $\{ \gamma_\mu, \gamma_* \} = 0$ for all $\gamma$ matrices, together with $\gamma_*^2 = \1$ and cyclicity of the trace.\footnote{We note that this in fact agrees with the prescription used by Bonora et al.\ for the parity-even part~\cite[footnote 4]{Bonora:2014qla}. What we do not agree with is the treatment of the parity-odd part, where one has to work from the beginning consistently in $n$ dimensions, as explained in Sec.~\ref{sec:expansion} after Eq.~\eqref{eq:fermion_prop}, and thus has to employ chiral projectors at every interaction vertex.} Applied to our situation, this means that instead of the result~\eqref{eq:1_spinorial} for the trace, we use
\begin{align}
\label{eq:1_spinorial_mod}
\tr\left( \mathcal{P}_+ \gamma^\tau \mathcal{P}_- \gamma^\mu \mathcal{P}_+ \gamma^\lambda \mathcal{P}_- \gamma^\rho \right) = 2 \left( \eta^{\tau\mu} \eta^{\lambda\rho} - \eta^{\tau\lambda} \eta^{\mu\rho} + \eta^{\tau\rho} \eta^{\mu\lambda} \right) - 2 \mathi \epsilon^{\tau\mu\lambda\rho} \eqend{,}
\end{align}
where the four-dimensional metrics $\bar\eta^{\mu\nu}$ in the parity-even part have been replaced by the $n$-dimensional ones $\eta^{\mu\nu}$. However, in the parity-odd part nothing has changed, and the totally antisymmetric symbol $\epsilon^{\tau\mu\lambda\rho}$ is still purely four-dimensional. Instead of eq.~\eqref{eq:VEV_2pf_integrated_b}, we obtain
\begin{splitequation}
\label{eq:VEV_2pf_integrated_mod}
\expect{ \Psi^{\mu\nu}(x) \Psi^{\rho\sigma}(y) } &= \int \frac{\mathcal{I}(p)}{n^2-1} \left[ 2 \Pi^{\mu(\nu}(p) \Pi^{\sigma)\rho}(p) - n \Pi^{\mu\rho}(p) \Pi^{\nu\sigma}(p) \right] \mathe^{\mathi p (x-y)} \dn p \eqend{,}
\end{splitequation}
which now has vanishing trace:
\begin{equation}
\label{eq:2pf_traces_mod}
\expect{ \Psi^{\mu\nu}(x) \Psi(y) } = \eta_{\mu\nu} \expect{ \Psi^{\mu\nu}(x) \Psi^{\rho\sigma}(y) } = 0 = \expect{ \Psi(x) \Psi^{\rho\sigma}(y) } \eqend{.}
\end{equation}
It follows that the regularized VEV of the stress tensor
\begin{splitequation}
\label{eq:VEV1_reg_mod}
\expect[reg]{ T^{\mu\nu}(x) }_{(1)} &= \frac{\mathi}{8 (n^2-1)} \iint \mathcal{I}(p) \Big[ \Pi^{\mu\nu}(p) \Pi^{\rho\sigma}(p) - (n-1) \Pi^{\rho(\mu}(p) \Pi^{\nu)\sigma}(p) \Big] \\
&\hspace{6em}\times \mathe^{\mathi p (x-y)} \dn p h_{\rho\sigma}(y) \total^n y
\end{splitequation}
is now both conserved and traceless as a consequence of the $n$-dimensional projector structure,
\begin{align}
\label{eq:1_reg_trace_mod}
\mathcal{T}^\text{reg}_{(1)}(x) = 0 = \mathcal{D}^\nu_{\text{reg},\,(1)}(x) \eqend{.}
\end{align}

Using again the \MS scheme, instead of eq.~\eqref{eq:VEV1_div} we thus obtain the divergent part of the stress tensor VEV
\begin{splitequation}
\label{eq:VEV1_div_mod}
\expect[div]{ T^{\mu\nu}(x) }_{(1)} &= - \frac{\diver}{120\cdot 16 \pi^2} \iint \Big[ \Pi^{\mu\nu}(p) \Pi^{\rho\sigma}(p) - 3 \Pi^{\rho(\mu}(p) \Pi^{\nu)\sigma}(p) \Big] \\
&\hspace{4em}\times \mathe^{\mathi p (x-y)} \dn p h_{\rho\sigma}(y) \total^n y \\
&= - \frac{\diver}{120\cdot 16 \pi^2} \left( 6 \nabla^2 R^{\mu\nu} - 2 \nabla^\mu \nabla^\nu R - g^{\mu\nu} \nabla^2 R \right)_{(1)}(x) \eqend{,}
\end{splitequation}
which only differs from eq.~\eqref{eq:VEV1_div} by the fact that all tensors are now taken in $n$ dimensions. Finally, the renormalized stress tensor VEV in the modified Breitenlohner--Maison scheme is given by
\begin{splitequation}
\label{eq:VEV1_ren_mod}
\expect[ren]{ T^{\mu\nu}(x) }_{(1)} &= \expect[reg]{ T^{\mu\nu}(x) }_{(1)} - \expect[div]{ T^{\mu\nu}(x) }_{(1)} \\
&= - \frac{1}{120\cdot 16 \pi^2} \iint \left[ \frac{12}{5} - \ln\left( \frac{p^2 - \mathi 0}{\mu^2} \right) \right] \\
&\hspace{4em}\times \left[ \Pi^{\mu\nu}(p) \Pi^{\rho\sigma}(p) - 3 \Pi^{\rho(\mu}(p) \Pi^{\nu)\sigma}(p) \right] \mathe^{\mathi p (x-y)} \dn[4] p h_{\rho\sigma}(y) \total^4 y \\
&\quad- \frac{1}{180\cdot 16 \pi^2} \iint \Pi^{\mu\nu}(p) \Pi^{\rho\sigma}(p) \mathe^{\mathi p (x-y)} \dn[4] p h_{\rho\sigma}(y) \total^4 y \eqend{.} \raisetag{2em}
\end{splitequation}
From the projectors, one sees that the first term is both conserved and traceless, while the second term is conserved but not traceless. Concretely, we obtain
\begin{equations}[eq:VEV1_tracediv_mod]
\mathcal{T}^\text{ren}_{(1)}(x) &= \frac{1}{60 (4\pi)^2} \left[ \nabla^2 R(x) \right]_{(1)} \eqend{,} \\
\mathcal{D}^\nu_{\text{ren},\, (1)}(x) &= 0 \eqend{.}
\end{equations}
This agrees with the result for a Majorana fermion or half the result for a Dirac fermion in dimensional regularization, see for example~\cite{Godazgar:2018boc}. Of course, as is well known, the term in the trace anomaly proportional to $\nabla^2 R$ can be canceled by adding a term proportional to $R^2$ to the action, such that we have agreement with the original Breitenlohner--Maison scheme up to the allowed freedom.

\subsection{The anomaly definition of Godazgar and Nicolai}

Lastly, we want to compare our computation of the anomalous trace with the definition of Godazgar and Nicolai~\cite{Godazgar:2018boc,Casarin:2018odz}. They postulate that the trace anomaly is given in general by
\begin{equation}
\mathcal{A} \equiv \lim_{n \to 4} \big( g^{\mu\nu} \expect{ T_{\mu\nu}(x) } - \expect{ g^{\mu\nu} T_{\mu\nu}(x) } \big) \eqend{,} 
\end{equation}
where the expectation values are the dimensionally regularized ones. This means that in the first term, with the metric outside the expectation value, the metric should be taken in $n = 4$ dimensions, while for the second term the $n$-dimensional metric enters. At first order, this results in
\begin{splitequation}
\label{eq:a1_godazgar_orig}
\mathcal{A}^\text{orig}_{(1)} &= \lim_{n \to 4}  \left( \bar{\eta}_{\mu\nu} - \eta_{\mu\nu} \right) \expect[reg]{ T^{\mu\nu}(x) }_{(1)} = - \lim_{n \to 4}  \hat{\eta}_{\mu\nu} \expect[reg]{ T^{\mu\nu}(x) }_{(1)} \\
&= \lim_{n \to 4} \frac{\mathi (n-4)^2}{16 (n^2-1)} \iint \mathcal{I}(p) p^2 \left[ \bar{\Pi}^{\rho\sigma}(p) + 3 \bar{\eta}^{\rho\sigma} p^2 \right] \mathe^{\mathi p (x-y)} \dn[4] p h_{\rho\sigma}(y) \total^4 y = 0
\end{splitequation}
with the regularized VEV~\eqref{eq:VEV1_reg} using the original Breitenlohner--Maison scheme, and
\begin{splitequation}
\label{eq:a1_godazgar_mod}
\mathcal{A}^\text{mod}_{(1)} &= - \lim_{n \to 4} \hat{\eta}_{\mu\nu} \expect[reg]{ T^{\mu\nu}(x) }_{(1)} \\
&= \lim_{n \to 4}  \frac{\mathi (n-4)}{8 (n^2-1)} \iint \mathcal{I}(p) p^2 \Pi^{\rho\sigma}(p) \, \mathe^{\mathi p (x-y)} \dn[4] p h_{\rho\sigma}(y) \total^4 y \\
&= \frac{1}{60 \cdot 16 \pi^2} \left[ \nabla^2 R \right]_{(1)} 
\end{splitequation}
with the regularized VEV~\eqref{eq:VEV1_reg_mod} using the modified Breitenlohner--Maison scheme.

We see that in both cases, a covariant result is obtained, and that the non-covariant finite counterterms that one needs in the original Breitenlohner--Maison scheme automatically cancel out. Nevertheless, the result between different schemes only agrees up to the usual freedom of adding finite covariant counterterms, as expected. We further note that since the renormalized stress tensor VEV is four-dimensional, we have $\hat\eta_{\mu\nu} \expect[ren]{ T^{\mu\nu}(x) }_{(1)} = 0$ and thus can also write
\begin{equation}
\label{eq:a1_godazgar_div}
\mathcal{A}_{(1)} = \lim_{n \to 4}  \hat{\eta}_{\mu\nu} \expect[div]{ T^{\mu\nu}(x) }_{(1)} \eqend{.}
\end{equation}
It follows that the difference between the results~\eqref{eq:a1_godazgar_orig} and~\eqref{eq:a1_godazgar_mod} comes about because of the different results for the divergent part in the two schemes, either using purely four-dimensional tensors in the original Breitenlohner--Maison scheme~\eqref{eq:VEV1_div} or using $n$-dimensional ones in the modified scheme~\eqref{eq:VEV1_div_mod}, see App.~\ref{app:counterterm}.

\section{The trace anomaly at \texorpdfstring{$\bigo{\kappa^2}$}{O(\textkappa2)}}
\label{sec:anomaly_k2}

At second order in $h$ the number of terms contributing to the stress tensor VEV increases substantially. In order to avoid a further proliferation of terms, and since the result for the parity-even part of the stress tensor VEV is not disputed, we will use the modified Breitenlohner--Maison scheme. We classify the contributions to $\expect{ T^{\mu\nu}(x) }_{(2)}$ into three categories, considering their structure in terms of the variables defined in~\eqref{eq:structures}: those which contain two $\Psi$'s, those with one $\Psi$ and one $J$, and finally those with three $\Psi$'s.
Using the computations of the previous section, we can easily give a result for the terms that fall either into the first or second category. After doing so, we will describe a way to perform the calculations for those in the third category, which we have done using the xAct tensor algebra suite~\cite{xact,xpert}, and give only the result for the divergence and trace.

\subsection{The \texorpdfstring{$\Psi^2$}{\textPsi2} contribution}
\label{sec:psi^2}

From the expansion of the stress tensor VEV~\eqref{eq:coefficients_VEV}, we see that there are contributions involving two $\Psi$'s both from $\expect{ S_{(2)} T^{\mu\nu}_{(0)}(x) }$ and $\expect{ S_{(1)} T^{\mu\nu}_{(1)}(x) }$. Using formula~\eqref{eq:VEV_2pf_integrated_mod} together with a reshuffling of the terms in order to render the expression readable, after some straightforward manipulations we arrive at the following expression for the regularized contribution to the VEV:
\begin{splitequation}
\expect[reg]{ T^{\mu\nu}(x) }_{(2),\,\Psi^2} &= \frac{1}{2} h_{\alpha\beta}(x) \eta^{\mu\nu} \expect[reg]{ T^{\alpha\beta}(x) }_{(1)} - \frac{3}{2} h^{(\mu}{}_\alpha(x) \expect[reg]{ T^{\nu)\alpha}(x) }_{(1)} \\
&\quad+ \frac{\mathi}{32 (n^2-1)} \iint \mathcal{I}(p) \left[ (n-1) \Pi^{\rho(\mu} \Pi^{\nu)\sigma} - \Pi^{\mu\nu} \Pi^{\rho\sigma} \right] \mathe^{\mathi p (x-y)} \dn p \\
&\hspace{8em}\times \left[ 3 h_{\alpha\rho}(y) h^\alpha{}_\sigma(y) - 2 h(y) h_{\rho\sigma}(y) \right] \total^n y \eqend{.} \raisetag{1.4em}
\end{splitequation}
Because of the projectors in the second term, this term does not contribute to the trace and divergence, and we obtain directly
\begin{equations}
\eta_{\mu\nu} \expect[reg]{ T^{\mu\nu}(x) }_{(2),\,\Psi^2} &= \frac{n-3}{2} h_{\alpha\beta}(x) \expect[reg]{ T^{\alpha\beta}(x) }_{(1)} \eqend{,} \label{eq:stress_trace_psi2} \\
\partial_\mu \expect[reg]{ T^{\mu\nu}(x) }_{(2),\Psi^2} &= \frac{1}{2} \partial_\mu \left[ h_{\alpha\beta}(x) \eta^{\mu\nu} \expect[reg]{ T^{\alpha\beta}(x) }_{(1)} - 3 h^{(\mu}{}_\alpha(x) \expect[reg]{ T^{\nu)\alpha}(x) }_{(1)} \right] \eqend{.} \label{eq:stress_div_psi2}
\end{equations}

\subsection{The \texorpdfstring{$\Psi J$}{\textPsi J} contribution}
\label{sec:psi_j}

The computation of the two-point function involving one $\Psi$ and one $J$ goes along the previous lines. One easily arrives at the following expression:
\begin{splitequation}
\expect{ \Psi^{\mu\nu}(x) J^{\alpha\beta\gamma}(y) } &= \mathi \tr\left( \gamma^\mu \mathcal{P}_+ \gamma^\lambda \mathcal{P}_- \gamma^{\alpha\beta\gamma} \mathcal{P}_+ \gamma^\tau \mathcal{P}_- \right) \\
&\quad\times \iint (2q+p)^\nu \frac{(p+q)_\tau}{(p+q)^2 - \mathi 0} \frac{q_\lambda}{q^2 - \mathi 0} \mathe^{\mathi p (x-y)} \dn p \dn q \eqend{,}
\end{splitequation}
where the integration over $q$ can again be done using the tensorial integrals~\eqref{eq:Ip} and results in
\begin{splitequation}
\label{eq:psij_VEV}
\expect{ \Psi^{\mu\nu}(x) J^{\alpha\beta\gamma}(y) } &= \frac{\mathi}{2 (n-1)} \tr\left( \gamma^\mu \mathcal{P}_+ \gamma^\lambda \mathcal{P}_- \gamma^{\alpha\beta\gamma} \mathcal{P}_+ \gamma^\tau \mathcal{P}_- \right) \\
&\quad\times \int \mathcal{I}(p) \, \delta^\nu_{[\tau} p_{\lambda]} p^2 \, \mathe^{\mathi p (x-y)} \dn p \eqend{.}
\end{splitequation}

For the $\gamma$ matrix trace we compute in the modified Breitenlohner--Maison scheme
\begin{splitequation}
\tr\left( \gamma^\mu \mathcal{P}_+ \gamma^\lambda \mathcal{P}_- \gamma^{\alpha\beta\gamma} \mathcal{P}_+ \gamma^\tau \mathcal{P}_- \right) &= 4 \left( \eta^{\alpha\tau} \eta^{\beta[\lambda} \eta^{\mu]\gamma} + \eta^{\alpha\mu} \eta^{\beta[\tau} \eta^{\lambda]\gamma} + \eta^{\alpha\lambda} \eta^{\beta[\mu} \eta^{\tau]\gamma} \right) \\
&\quad+ \frac{1}{2} \tr\left( \gamma^\mu \gamma_* \bar\gamma^\lambda \bar\gamma^{\alpha\beta\gamma} \bar\gamma^\tau \right)
\end{splitequation}
using formula~\eqref{eq:gamma_sandwich_projector}, where the parity-odd contribution is given by
\begin{splitequation}
\label{eq:spinorial_Jpsi_odd}
\tr \left( \gamma^\mu \gamma_* \bar\gamma^\lambda \bar\gamma^{\alpha\beta\gamma} \bar\gamma^\tau \right) &= 12 \mathi \left( \epsilon^{\lambda\tau[\alpha\beta} \bar\eta^{\gamma]\mu} + \epsilon^{\mu\lambda[\alpha\beta} \bar\eta^{\gamma]\tau} + \epsilon^{\mu\tau[\alpha\beta} \bar\eta^{\gamma]\lambda} \right) \\
&\quad+ 4 \mathi \left( \epsilon^{\alpha\beta\gamma\mu} \bar\eta^{\lambda\tau} + \epsilon^{\alpha\beta\gamma\lambda} \bar\eta^{\tau\mu} - \epsilon^{\alpha\beta\gamma\tau} \bar\eta^{\lambda\mu} \right) \eqend{.}
\end{splitequation}
However, unlike the parity-odd contribution to the trace appearing in the contribution of the VEV of two $\Psi$'s~\eqref{eq:1_spinorial}, this expression is not unique because of dimension-dependent identities~\cite{Edgar:2001vv}. Namely, for purely four-dimensional tensors, antisymmetri\-zation over five or more indices obviously gives a vanishing result. Therefore, adding the vanishing expression
\begin{equation}
- 20 \mathi \left( \epsilon^{[\alpha\beta\gamma\mu} \bar\eta^{\tau]\lambda} + \epsilon^{[\alpha\beta\gamma\mu} \bar\eta^{\lambda]\tau} + \epsilon^{[\alpha\beta\gamma\lambda} \bar\eta^{\tau]\mu} \right)
\end{equation}
to the result~\eqref{eq:spinorial_Jpsi_odd} we obtain the simpler form
\begin{equation}
\label{eq:spinorial_Jpsi_odd_short}
\tr \left( \gamma^\mu \gamma_* \bar\gamma^\lambda \bar\gamma^{\alpha\beta\gamma} \bar\gamma^\tau \right) = 4 \mathi \left( 2 \epsilon^{\alpha\beta\gamma(\tau} \bar\eta^{\lambda)\mu} - \epsilon^{\alpha\beta\gamma\mu} \bar\eta^{\lambda\tau} \right) \eqend{.}
\end{equation}

Since the $p$ integral of the VEV~\eqref{eq:psij_VEV} is antisymmetric and the trace~\eqref{eq:spinorial_Jpsi_odd_short} symmetric in $\tau\lambda$, there is no parity-odd contribution to the VEV of one $\Psi$ and one $J$. The parity-even contribution reads
\begin{equation}
\label{eq:psij_VEV_b}
\expect{ \Psi^{\mu\nu}(x) J^{\alpha\beta\gamma}(y) } = - \frac{6 \mathi}{(n-1)} \int \mathcal{I}(p) \, \eta^{\mu[\alpha} p^\beta \eta^{\gamma]\nu} p^2 \, \mathe^{\mathi p (x-y)} \dn p \eqend{,}
\end{equation}
and is antisymmetric in $\mu\nu$. However, the contribution of~\eqref{eq:psij_VEV_b} to the stress tensor VEV only involves the symmetric part $\Psi^{(\mu\nu)}$ at second order: from eq.~\eqref{eq:coefficients_VEV} we see that there is a contribution from $\expect{ S_{(2)} T^{\mu\nu}_{(0)}(x) }$, where $S_{(2)}$~\eqref{eq:actionexpansion_s2} involves $J^{\alpha\beta\gamma}$ and $T^{\mu\nu}_{(0)}$~\eqref{eq:emtensorexpansion_0} the symmetric part $\Psi^{(\alpha\beta)}$, while for the contribution from $\expect{ S_{(1)} T^{\mu\nu}_{(1)}(x) }$ the action $S_{(1)}$~\eqref{eq:actionexpansion_s1} involves the symmetric part $\Psi^{(\alpha\beta)}$ and the stress tensor $T^{\mu\nu}_{(1)}(x)$~\eqref{eq:emtensorexpansion_1} involves $J^{\alpha\beta\gamma}$.\footnote{The first possible contribution of the antisymmetric part $\Psi^{[\mu\nu]}$ is at third order from $\expect{ S_{(2)} T^{\mu\nu}_{(1)}(x) }$.}

We conclude that the contribution from the VEV of one $\Psi$ and one $J$ to the stress tensor vanishes:
\begin{align}
\expect[reg]{ T^{\mu\nu}(x) }_{(2),\,\Psi J} = 0 \label{eq:stress_psij} \eqend{.}
\end{align}

\subsection{The \texorpdfstring{$\Psi^3$}{\textPsi3} contribution}
\label{sec:psi^3}

We arrive therefore to the most challenging computation of this article, the contribution of three $\Psi$'s in the term $\expect{ S_{(1)}^2 T^{\mu\nu}_{(0)}(x) }$~\eqref{eq:coefficients_VEV}. Even if the computations involved are not necessarily conceptually complex, they are rather lengthy ($\sim 10^4-10^5$ terms), and even with computer algebra take quite some time to finish. Let us begin by giving a concrete expression for the three-point contribution:
\begin{splitequation}
\label{eq:psi3_integral}
&\expect{ \Psi^{\mu\nu}(x) \Psi^{\alpha\beta}(y) \Psi^{\sigma\rho}(z) } \\
&= - \mathi \, \left\lbrace \tr\left[ \gamma^\mu \mathcal{P}_+ \gamma^\tau \mathcal{P}_- \gamma^\sigma \mathcal{P}_+ \gamma^\delta \mathcal{P}_- \gamma^\alpha \mathcal{P}_+ \gamma^\lambda \mathcal{P}_- \right] + \tr\left[ \gamma^\mu \mathcal{P}_+ \gamma^\lambda \mathcal{P}_- \gamma^\alpha \mathcal{P}_+ \gamma^\delta \mathcal{P}_- \gamma^\sigma \mathcal{P}_+ \gamma^\tau \mathcal{P}_- \right] \right\rbrace \\
&\hspace{3em}\times \iiint \frac{(q-k)_\tau q_\delta (q-p)_\lambda (2q-k)^\rho (2q-p)^\beta (2q-p-k)^\nu}{[ (q-k)^2 - \mathi 0 ] [ (q-p)^2 - \mathi 0 ] [ q^2 - \mathi 0 ]} \\
&\hspace{10em}\times \mathe^{-\mathi p (x-y)} \mathe^{\mathi k (x-z)} \dn p \dn q \dn k \eqend{.} \raisetag{2em}
\end{splitequation}
We first compute the $\gamma$ matrix traces using the modified Breitenlohner--Maison scheme, which results in
\begin{splitequation}
\label{eq:psi3_gammatrace}
&\tr\left[ \gamma^\mu \mathcal{P}_+ \gamma^\tau \mathcal{P}_- \gamma^\sigma \mathcal{P}_+ \gamma^\delta \mathcal{P}_- \gamma^\alpha \mathcal{P}_+ \gamma^\lambda \mathcal{P}_- \right] + \tr\left[ \gamma^\mu \mathcal{P}_+ \gamma^\lambda \mathcal{P}_- \gamma^\alpha \mathcal{P}_+ \gamma^\delta \mathcal{P}_- \gamma^\sigma \mathcal{P}_+ \gamma^\tau \mathcal{P}_- \right] \\
&\quad= 4 \Bigl[ 2 \eta^{\alpha[\lambda} \left( 2 \eta^{\mu][\tau} \eta^{\sigma]\delta} + \eta^{\mu]\delta} \eta^{\sigma\tau} \right) + 2 \eta^{\alpha[\delta} \left( 2 \eta^{\sigma][\lambda} \eta^{\mu]\tau} + \eta^{\sigma]\tau} \eta^{\lambda\mu} \right) \\
&\hspace{4em}+ \eta^{\alpha\tau} \left( 2 \eta^{\delta[\sigma} \eta^{\mu]\lambda} + \eta^{\delta\lambda} \eta^{\mu\sigma} \right) \Bigr] + 4 \mathi \epsilon^{[\delta\lambda\sigma\tau} \bar\eta^{\alpha]\mu} \eqend{.}
\end{splitequation}
At first sight, one could think that we have found one additional contribution of odd parity. However, the corresponding term involves antisymmetrization over five indices, and since the involved tensors are purely four-dimensional, this contribution again vanishes.

For the parity-even contribution, we first note that by taking either a trace or divergence of eq.~\eqref{eq:psi3_integral} and performing the contractions with the metrics of eq.~\eqref{eq:psi3_gammatrace}, the integrals reduce to the tensorial integrals~\eqref{eq:Ip}, of argument $p$, $k$ or $p-k$, and can be expressed in terms of the two-point function~\eqref{eq:VEV_2pf_integrated_mod}. After a laborious but straightforward computation, we find that the regularized contributions to the trace and the divergence of the stress tensor are given by
\begin{equations}
\eta_{\mu\nu} \expect[reg]{ T^{\mu\nu}(x) }_{(2),\,\Psi^3} &= - \frac{n (n-1)}{8} h_{\rho\sigma}(x) \expect[reg]{ T^{\rho\sigma}(x) }_{(1)} \eqend{,} \label{eq:stress_trace_psi3} \\
\begin{split}
\partial_\mu \expect[reg]{ T^{\mu\nu}(x) }_{(2),\Psi^3} &= - \frac{1}{2} \partial_\rho h(x) \expect[reg]{ T^{\nu\rho}(x) }_{(1)} + \frac{3}{4} \partial^\rho \left[ h_{\rho\sigma}(x) \expect[reg]{ T^{\nu\sigma}(x) }_{(1)} \right] \\
&\quad- \frac{1}{4} \partial_\rho h_\sigma{}^\nu \expect[reg]{ T^{\rho\sigma}(x) }_{(1)} - \frac{1}{2} h_{\rho\sigma}(x) \partial^\nu \expect[reg]{ T^{\rho\sigma}(x) }_{(1)} \eqend{.} \label{eq:stress_div_psi3}
\end{split}
\end{equations}
It remains to compute the divergent part of eq.~\eqref{eq:psi3_integral} in order to obtain the trace and divergence of the renormalized stress tensor.

\subsubsection{Isolating the divergences in the \texorpdfstring{$\Psi^3$}{O(\textPsi3)} contribution}
\label{sec:psi^3_div}

As previously said, there is a large number of terms involved in the computation, which renders it somewhat unfeasible. The experience gained in the preceding sections invites us to define the two-argument tensorial integrals
\begin{align}
\label{eq:Ipk}
\mathcal{I}^{\mu_1 \cdots \mu_m}(p,k) \equiv \int \frac{q^{\mu_1} \cdots q^{\mu_m}}{( q^2 - \mathi 0 ) [ (q-k)^2 - \mathi 0 ] [ (q-p)^2 - \mathi 0 ]} \dn q \eqend{,}
\end{align}
which one could evaluate using the results of Godazgar and Nicolai~\cite{Godazgar:2018boc}, and extract the divergent part. However, since their approach leads to spurious kinematical singularities (terms of the form $\left[ p^2 k^2 - (pk)^2 \right]^{-1}$, which only can cancel at the very end), it is more useful for our purposes to proceed in a different way.

As already noted in~\cite{Godazgar:2018boc}, if one is only interested in the divergent part one can proceed as follows: by power counting, $\mathcal{I}(p,k)$ and $\mathcal{I}^\mu(p,k)$ are finite in $n = 4$ dimensions, while $\mathcal{I}^{\mu\nu}(p,k)$ is logarithmically divergent. The divergence of $\mathcal{I}^{\mu\nu}(p,k)$ must therefore be proportional to $\eta^{\mu\nu}$, and (again by power counting) the divergences of $\mathcal{I}^{\mu_1 \cdots \mu_m}(p,k)$ must be proportional to a polynomial in $p$ and $k$ of order $m-2$. We then use the important property (which was already mentioned) that the partial trace of $\mathcal{I}^{\mu \cdots}(p,k)$ is given by combinations of the one-argument tensorial integrals defined in eq.~\eqref{eq:Ip}:
\begin{splitequation}
\label{eq:Ipk_trace}
\eta_{\mu\nu} \, \mathcal{I}^{\mu\nu\rho\cdots}(p,k) &= \int \frac{q^\rho \cdots}{\left[ (q-p)^2 - \mathi0 \right]^2 \left[ (q-k)^2 - \mathi0 \right]^2} \dn q \\
&= \mathcal{I}^{\rho\cdots}(p-k) + p^\rho \, \mathcal{I}^{\cdots}(p-k) + \cdots \eqend{,}
\end{splitequation}
where we have shifted the integration variable $q \to q+p$. Together with the manifest symmetries $\mathcal{I}^{\mu_1 \cdots \mu_m}(p,k) = \mathcal{I}^{(\mu_1 \cdots \mu_m)}(p,k) = \mathcal{I}^{\mu_1 \cdots \mu_m}(k,p)$ which can be read off from~\eqref{eq:Ipk}, this relation suffices to uniquely determine the divergent part, which is seen to be a combination of the one-argument tensorial integrals~\eqref{eq:Ip}.

However, we would like to compute the full expectation value which also includes additional finite parts, and for which we need to evaluate the two-argument tensorial integrals~\eqref{eq:Ipk} explicitly. From eq.~\eqref{eq:Ipk_trace}, it follows that these additional finite parts are traceless. To compute the integrals~\eqref{eq:Ipk}, we introduce Feynman parameters~\cite{smirnov} to combine the denominators and obtain after the shift $q \to q + y p + x k$ that
\begin{align}
\mathcal{I}^{\mu_1 \cdots \mu_m}(p,k) = 2 \int_0^1 \int_0^{1-y} \int \frac{(q+yp+xk)^{\mu_1} \cdots}{\left( q^2 + M_E - \mathi 0 \right)^3} \dn q \total x \total y \eqend{,}
\end{align}
where
\begin{align}
\label{eq:me_def}
M_E \equiv y (1-y) p^2 + x (1-x) k^2 - 2 x y (p\cdot k) \eqend{.}
\end{align}
The integral over $q$ can now be done in the standard way~\cite{smirnov} by first reducing them to scalar ones. In particular, integrals that contain an odd number of $q$'s vanish by the symmetry $q \to -q$, and the tensorial structure of the integrals with an even number of $q$'s is by rotational invariance obtained with the replacements
\begin{equations}
q^\mu q^\nu &\to \frac{1}{n} \eta^{\mu\nu} q^2 \eqend{,} \\
q^\mu q^\nu q^\rho q^\sigma &\to \frac{3}{n (n+2)} \eta^{(\mu\nu} \eta^{\rho\sigma)} q^4 \eqend{,} \\
q^\mu q^\nu q^\rho q^\sigma q^\alpha q^\beta &\to \frac{15}{n (n+2) (n+4)} \eta^{(\mu\nu} \eta^{\rho\sigma} \eta^{\alpha\beta)} q^6 \eqend{.}
\end{equations}
However, any power of $q^2$ cancels the one in the denominator of~\eqref{eq:Ipk}, such that a one-argument integrals results that contributes to the divergent part. It is thus clear that the additional finite part is given by the traceless part (denoted by a subscript ``tr'')
\begin{splitequation}
\trlss{ \mathcal{I}^{\mu \cdots}(p,k) } = 2 \int_0^1 \int_0^{1-y} \int \frac{\trlss{ (yp+xk)^\mu \cdots }}{\left( q^2 + M_E - \mathi 0 \right)^3} \dn q \total x \total y \eqend{,}
\end{splitequation}
which by power counting is indeed seen to be finite in $n = 4$ dimensions. The $q$ integral can then be done using~\cite{smirnov}
\begin{equation}
\int \frac{1}{\left( q^2 + M_E - \mathi 0 \right)^a} \frac{\total^n q}{(2\pi)^n} = \frac{\mathi}{(4\pi)^\frac{n}{2}} \frac{\Gamma\left( a - \frac{n}{2} \right)}{\Gamma(a)} \left( M_E - \mathi 0 \right)^{\frac{n}{2}-a} \eqend{,}
\end{equation}
and the limit $n \to 4$ can be taken, and we see that the result can be expressed using the parameter integrals
\begin{equation}
\label{eq:F}
F_{ab}(k,p) \equiv \int_0^1 \int_0^{1-y} \frac{x^a y^b}{M_E - \mathi 0} \total x \total y \eqend{.}
\end{equation}
To our knowledge, these integrals were first introduced in~\cite{Adler:1969gk} in the study of axial anomalies (denoted there by $I_{st}$~\cite[eq.~(18)]{Adler:1969gk}), and we note that they satisfy several relations which are given (and proven) in App.~\ref{app:FG}. Taking all together, for the first three two-argument integrals we obtain
\begin{equations}
\mathcal{I}(p,k) &= \frac{\mathi}{16 \pi^2} F_{00}(k,p) \eqend{,} \\
\mathcal{I}^\mu(p,k) &= \frac{\mathi}{16 \pi^2} \left[ F_{10}(k,p) k^\mu + F_{01}(k,p) p^\mu \right] \eqend{,} \\
\begin{split}
\mathcal{I}^{\mu\nu}(p,k) &= \frac{\eta^{\mu\nu}}{n} \, \mathcal{I}(p-k) + \frac{\mathi}{16 \pi^2} \trlss{ F_{20}(k,p) k^\mu k^\nu + 2 F_{11}(k,p) k^{(\mu} p^{\nu)} + F_{02}(k,p) p^\mu p^\nu } \eqend{,}
\end{split}
\end{equations}
and the remaining integrals relevant for this work are given in App.~\ref{app:integrals_2}. In this way, the computation is reduced to around $10^3$ terms, and all kinematical singularities are avoided. Since the final result is not really inspiring, we won't include its explicit expression in the text; the interested reader can consult the ancillary Mathematica file.

\subsection{The complete \texorpdfstring{$\bigo{\kappa^2}$}{O(\textkappa2)} contribution}
\label{sec:trace_k2}

At this point we have all the necessary ingredients to compute the full second-order contribution to the trace and divergence of the stress tensor VEV. Remembering to include the terms in eqs.~\eqref{eq:tr_orderh2} and~\eqref{eq:div_orderh2} explicitly proportional to $h$, we get a regularized expression which is both traceless and divergence-free at second order:
\begin{equations}[eq:tracediv_k2]
\begin{split}
\mathcal{T}^\text{reg}{}_{(2)}(x) &= \eta_{\mu\nu} \left[ \expect[reg]{ T^{\mu\nu}(x) }_{(2),\,\Psi^2} + \expect[reg]{ T^{\mu\nu}(x) }_{(2),\,\Psi J} + \expect[reg]{ T^{\mu\nu}(x) }_{(2),\,\Psi^3} \right] + h_{\mu\nu} \expect[reg]{ T^{\mu\nu}(x) }_{(1)} \\
&= 0 \eqend{,} \raisetag{1.8em}
\end{split} \\
\begin{split}
\mathcal{D}^\nu_{\text{reg},\,(2)}(x) &= \partial_\mu \left[ \expect[reg]{ T^{\mu\nu}(x) }_{(2),\,\Psi^2} + \expect[reg]{ T^{\mu\nu}(x) }_{(2),\,\Psi J} + \expect[reg]{ T^{\mu\nu}(x) }_{(2),\,\Psi^3} \right] \\
&\qquad+ \frac{1}{2} \left[ 2 \partial_\rho h^\nu{}_\mu(x) - \partial^\nu h_{\rho\mu}(x) + \delta^\nu_\mu \partial_\rho h(x) \right] \expect[reg]{ T^{\mu\rho}(x) }_{(1)} = 0 \eqend{,}
\end{split}
\end{equations}
where we have used the results~\eqref{eq:stress_trace_psi2}, \eqref{eq:stress_psij} and~\eqref{eq:stress_trace_psi3} for the trace and~\eqref{eq:stress_div_psi2}, \eqref{eq:stress_psij} and~\eqref{eq:stress_div_psi3} for the divergence.

As done at linear order in $h$, we renormalize using the \MS scheme, subtracting the contributions proportional to the divergent part $\diver$ of the scalar integral~\eqref{eq:scalar_integral}. Identifying once more the various terms as expansions of geometrical quantities by comparing with the results of App.~\ref{app:geometric_expansion}, we obtain
\begin{splitequation}
\label{eq:VEV2_div}
\expect[div]{ T^{\mu\nu}(x) }_{(2)} &= \frac{\diver}{1440 (4\pi)^2} \Big[ 8 g^{\mu\nu} R_{\alpha\beta} R^{\alpha\beta} + 56 R^{\mu\alpha} R^\nu_\alpha + 20 R^{\mu\nu} R - 5 g^{\mu\nu} R^2 \\
&\hspace{6em}+ 7 g^{\mu\nu} R_{\alpha\beta\gamma\delta} R^{\alpha\beta\gamma\delta} - 88 R_{\alpha\beta} R^{\mu\alpha\nu\beta} - 28 R^{\mu\alpha\beta\gamma} R^\nu{}_{\alpha\beta\gamma} \\
&\hspace{6em}- 72 \nabla^2 R^{\mu\nu} + 12 g^{\mu\nu} \nabla^2 R + 24 \nabla^\mu \nabla^\nu R \Big]_{(2)} \eqend{,}
\end{splitequation}
and the renormalized stress tensor VEV
\begin{equation}
\expect[ren]{ T^{\mu\nu}(x) }_{(2)} = \expect[reg]{ T^{\mu\nu}(x) }_{(2)} - \expect[div]{ T^{\mu\nu}(x) }_{(2)} \eqend{.}
\end{equation}
From this expression and eq.~\eqref{eq:tracediv_k2}, the trace and divergence of the renormalized stress tensor follow immediately. In terms of the Weyl tensor $C_{\mu\nu\rho\sigma}$ and the four-dimensional Euler density $\mathcal{E}_4$, which in four dimensions satisfy
\begin{equations}
C^{\mu\nu\rho\sigma} C_{\mu\nu\rho\sigma} &= R^{\mu\nu\rho\sigma} R_{\mu\nu\rho\sigma} - 2 R^{\mu\nu} R_{\mu\nu} + \frac{1}{3} R^2 \eqend{,} \\
\mathcal{E}_{4} &= R^{\mu\nu\rho\sigma} R_{\mu\nu\rho\sigma} - 4 R^{\mu\nu} R_{\mu\nu} +  R^2 \eqend{,}
\end{equations}
our results are exactly half of the trace anomaly for the Dirac spinor~\cite{Godazgar:2018boc}
\begin{equations}
\mathcal{T}^\text{ren}_{(2)}(x) &= \frac{1}{16 \cdot 45 (4\pi)^2} \left( - 11 \mathcal{E}_4 + 18 C^{\mu\nu\rho\sigma} C_{\mu\nu\rho\sigma} + 12 \nabla^2 R \right)_{(2)} \eqend{,} \\
\mathcal{D}^{\nu}_{\text{ren},\,(2)}(x) &= 0 \eqend{.}
\end{equations}

\section{Conclusions}
\label{sec:conclusions}

We have computed the renormalized stress tensor expectation value for chiral fermions using dimensional regularization, up to second order in perturbations around flat spacetime. Employing the Breitenlohner--Maison scheme~\cite{Breitenlohner:1977hr} for the treatment of the chiral matrix $\gamma_*$, we did not find any parity-odd contributions to the expectation value. Therefore, also the trace anomaly does not contain any parity-odd term proportional to the Pontryagin density, and the result for the trace anomaly is half the one of a Dirac fermion, as also confirmed using other methods~\cite{Bastianelli:2016nuf,Bastianelli:2019zrq,Frob:2019dgf}. This result reaffirms the validity of the equivalence principle for the coupling of matter to gravity in the present case, since there is no difference between left- or right-handed fermions.

For our result, the use of dimension-dependent identities~\cite{Edgar:2001vv} was crucial. As emphasized in~\cite{Korner:1991sx}, for $\gamma$ matrix traces containing one chiral matrix $\gamma_*$ and up to four other $\gamma$ matrices (in four dimensions), any prescription gives the correct result. Only for traces containing at least six other $\gamma$ matrices various features of the concrete prescription (such as the four-dimensionality in the Breitenlohner--Maison scheme, or the non-cyclicity of the trace for the prescription of Kreimer et al.~\cite{Kreimer:1989ke,Korner:1991sx}) become relevant for practical computations. This can also be seen in our computation, where at linear order in $h$ the parity-odd contribution to the $\gamma$ matrix trace~\eqref{eq:1_spinorial} is unique (proportional to the totally antisymmetric symbol $\epsilon$), while at second order this contribution can take various superficially different forms such as~\eqref{eq:spinorial_Jpsi_odd} or \eqref{eq:spinorial_Jpsi_odd_short}. However, strictly four-dimensional identities (stemming from the vanishing of terms that are antisymmetrized in five indices) show that all these forms are equal, since in the Breitenlohner--Maison scheme all involved tensors are purely four-dimensional and these identities are applicable.

As an extension of the present computation, we plan to reproduce in a future publication the Kimura--Delbourgo--Salam anomaly~\cite{Kimura:1969wj,Delbourgo:1972xb}, the anomalous divergence of the axial current $j^\mu_* = \mathi \bar{\psi} \gamma^\mu \gamma_* \psi$ in a gravitational field, by a full computation of its expectation value in the Breitenlohner--Maison scheme. The anomaly takes the form
\begin{equation}
\label{eq:kds_anomaly}
\nabla_\mu \expect{ j^\mu_* } = \frac{1}{\sqrt{-g}} \partial_\mu \left( \sqrt{-g} \expect{ j^\mu } \right) = \frac{1}{384 \pi^2} \varepsilon^{\mu\nu\rho\sigma} R_{\mu\nu\alpha\beta} R_{\rho\sigma}{}^{\alpha\beta} \eqend{,}
\end{equation}
where $\varepsilon^{\mu\nu\rho\sigma} = \epsilon^{\mu\nu\rho\sigma}/\sqrt{-g}$ is the Levi-Civita tensor. However, in our case the following happens: first, for a left-handed Weyl fermion we have $\gamma_* \psi = \psi$~\eqref{eq:weyl_fermion_def} and thus $j^\mu_* = j^\mu$ with the gauge current $j^\mu = \mathi \bar{\psi} \gamma^\mu \psi$.\footnote{This is of course the reason why the axial/chiral anomalies, which are associated with a global symmetry and thus physically harmless for theories involving Dirac fermions, become disastrous for chiral fermions, because then they destroy the gauge symmetry of the theory (whose interactions must involve conserved currents).} Second, the Pontryagin density is a total derivative, and to second order in the expansion around flat spacetime we have using the expansions of App.~\ref{app:geometric_expansion}
\begin{equation}
\epsilon^{\mu\nu\rho\sigma} R_{\mu\nu\alpha\beta} R_{\rho\sigma}{}^{\alpha\beta} = \partial_\mu \left( 4 \kappa^2 \epsilon^{\mu\nu\rho\sigma} \partial_{[\alpha} h_{\beta]\nu} \partial^\alpha \partial_\rho h_\sigma{}^\beta \right) + \bigo{\kappa^3} \eqend{.}
\end{equation}
It thus seems that one could cancel this anomaly by adding
\begin{equation}
- \frac{1}{96 \pi^2} \kappa^2 A_\mu \epsilon^{\mu\nu\rho\sigma} \partial_{[\alpha} h_{\beta]\nu} \partial^\alpha \partial_\rho h_\sigma{}^\beta + \bigo{\kappa^3}
\end{equation}
to the action, changing in this way the current $j^\mu = j^\mu_*$, since in the Breitenlohner--Maison scheme non-covariant counterterms are allowed and are actually necessary: see eq.~\eqref{eq:VEV1_finite_ct}. However, such a term changes also the stress tensor, and in particular its trace
\begin{equation}
\mathcal{T} \to \mathcal{T} + \frac{\kappa}{192 \pi^2} \partial_\alpha \left( F_{\rho\sigma} \epsilon^{\rho\sigma\mu\nu} \partial_\nu h_\mu{}^\alpha \right) + \bigo{\kappa^2}
\end{equation}
and divergence
\begin{equation}
\mathcal{D}^\nu \to \mathcal{D}^\nu + \frac{\kappa}{192 \pi^2} \partial_\alpha \left( F_{\rho\sigma} \epsilon^{\rho\sigma\mu\beta} \partial^{[\nu} \partial_\mu h_\beta{}^{\alpha]} \right) + \bigo{\kappa^2} \eqend{.}
\end{equation}
Therefore, one would have to compute also the contribution from a background gauge field to the stress tensor expectation value. Demanding that the stress tensor be conserved and have a gauge-invariant trace anomaly will then fix unambiguously also the Kimura--Delbourgo--Salam anomaly~\eqref{eq:kds_anomaly}.

\begin{acknowledgments}
SAF is grateful to G. Gori and the Institut f{\"u}r Theoretische Physik, Heidelberg, for their kind hospitality.
This work has been funded by the Deutsche Forschungsgemeinschaft (DFG, German Research Foundation) --- project nos. 415803368 and 406116891 within the Research Training Group RTG 2522/1.
SAF acknowledges support by UNLP, under project grant X909 and ``Subsidio a J{\'o}venes Investigadores 2019''.
We thank F. Schaposnik and F. Bastianelli for discussions.
\end{acknowledgments}

\appendix

\section{The tensorial integrals \texorpdfstring{$\mathcal{I}(p)$}{I(p)}}
\label{app:integrals_1}

For completeness, we give here the explicit expressions for the tensorial integrals defined in eq.~\eqref{eq:Ip} that are relevant for the present computation. More information can be found, e.g.,  in~\cite{smirnov}; the result for the scalar integral $\mathcal{I}(p)$ was given in eq.~\eqref{eq:scalar_integral}.
\begin{equations}
\mathcal{I}^\mu(p) &= - \frac{1}{2} p^\mu \, \mathcal{I}(p) \eqend{,} \\
\mathcal{I}^{\mu\nu}(p) &= \frac{1}{4 (n-1)} \left[ n p^\mu p^\nu - \eta^{\mu\nu} p^2 \right] \mathcal{I}(p) \eqend{,} \\
\mathcal{I}^{\mu\nu\rho}(p) &= - \frac{1}{8 (n-1)} \left[ (n+2) p^\mu p^\nu p^\rho - 3 p^2 \eta^{(\mu\nu} p^{\rho)} \right] \mathcal{I}(p) \eqend{,} \\
\begin{split}
\mathcal{I}^{\mu\nu\rho\sigma}(p) &= \frac{1}{16 (n-1) (n+1)} \Big[ (n+4) (n+2) p^\mu p^\nu p^\rho p^\sigma 
\\
&\hspace{3cm}- 6 (n+2) p^2 \eta^{(\mu\nu} p^\rho p^{\sigma)} + 3 (p^2)^2 \eta^{(\mu\nu} \eta^{\rho\sigma)} \Big] \mathcal{I}(p) \eqend{,}
\end{split} \\
\begin{split}
\mathcal{I}^{\mu\nu\rho\sigma\alpha}(p) &= - \frac{1}{32 (n-1) (n+1) (n+2)} \Big[ (n+6) (n+4) (n+2) p^\mu p^\nu p^\rho p^\sigma p^\alpha \\
&\quad- 10 (n+4) (n+2) p^2 \eta^{(\mu\nu} p^\rho p^\sigma p^{\alpha)} + 15 (n+2) (p^2)^2 \eta^{(\mu\nu} \eta^{\rho\sigma} p^{\alpha)} \Big] \mathcal{I}(p) \eqend{.}
\end{split}
\end{equations}
We note that these expressions are uniquely determined by the tracelessness condition $\eta_{\mu\nu} \, \mathcal{I}^{\mu\nu \cdots}(p) = 0$, the symmetry $\mathcal{I}^{\mu_1 \cdots \mu_m}(p) = \mathcal{I}^{(\mu_1 \cdots \mu_m)}(p)$, and the contraction
\begin{equation}
2 p_\mu \, \mathcal{I}^{\mu\nu \cdots}(p) = - p^2 \, \mathcal{I}^{\nu \cdots}(p) \eqend{,}
\end{equation}
all of which follow directly from the definition~\eqref{eq:Ip} by using the fact that scaleless integrals (with only a single power in the denominator) vanish in dimensional regularization~\cite{Leibbrandt:1975dj}. In fact, one can easily give the general result~\cite{smirnov}:
\begin{equation}
\mathcal{I}^{\mu_1 \cdots \mu_m}(p) = (-1)^m \frac{\Gamma(n-1) \Gamma\left( \frac{n-2}{2} + m \right)}{2 \Gamma(n-2+m) \Gamma\left( \frac{n}{2} \right)} \trlss{ p^{\mu_1} \cdots p^{\mu_m} } \, \mathcal{I}(p) \eqend{,}
\end{equation}
where the subscript ``tr'' denotes the traceless part, given by
\begin{splitequation}
\label{eq:traceless}
\trlss{ p^{\mu_1} \cdots p^{\mu_m} } &= \sum_{r=0}^{[m/2]} \frac{(-1)^r m!}{4^r r! (m-2r)!} \frac{\Gamma\left( \frac{n-2}{2} + m - r \right)}{\Gamma\left( \frac{n-2}{2} + m \right)} \\
&\qquad\times (p^2)^r \eta^{(\mu_1\mu_2} \cdots \eta^{\mu_{2r-1}\mu_{2r}} p^{\mu_{2r+1}} \cdots p^{\mu_m)} \eqend{.}
\end{splitequation}

\section{On the existence of suitable counterterms}
\label{app:counterterm}

Summing the results~\eqref{eq:VEV1_div_mod} at first order and~\eqref{eq:VEV2_div} at second order, we have
\begin{splitequation}
\label{eq:VEV_div}
\expect[div]{ T^{\mu\nu}(x) } &= \frac{\diver}{1440 (4\pi)^2} \Big[ 8 g^{\mu\nu} R_{\alpha\beta} R^{\alpha\beta} + 56 R^{\mu\alpha} R^\nu_\alpha + 20 R^{\mu\nu} R - 5 g^{\mu\nu} R^2 \\
&\hspace{6em}+ 7 g^{\mu\nu} R_{\alpha\beta\gamma\delta} R^{\alpha\beta\gamma\delta} - 88 R_{\alpha\beta} R^{\mu\alpha\nu\beta} - 28 R^{\mu\alpha\beta\gamma} R^\nu{}_{\alpha\beta\gamma} \\
&\hspace{6em}- 72 \nabla^2 R^{\mu\nu} + 12 g^{\mu\nu} \nabla^2 R + 24 \nabla^\mu \nabla^\nu R \Big] + \bigo{\kappa^3} \eqend{,}
\end{splitequation}
and want to show that this can be canceled by a local counterterm.

With the variations~\cite{xpert}
\begin{equations}
\delta \left( R^{\mu\nu\rho\sigma} R_{\mu\nu\rho\sigma} \sqrt{-g} \right) &= - \sqrt{-g} \left[ 2 R^{\mu\alpha\beta\gamma} R^\nu{}_{\alpha\beta\gamma} - \frac{1}{2} R^{\alpha\beta\gamma\delta} R_{\alpha\beta\gamma\delta} g^{\mu\nu} + 4 R^{\rho\mu\sigma\nu} \nabla_\rho \nabla_\sigma \right] \delta g_{\mu\nu} \eqend{,} \raisetag{2em} \\
\begin{split}
\delta \left( R^{\mu\nu} R_{\mu\nu} \sqrt{-g} \right) &= - \sqrt{-g} \Big[ 2 R^{\mu\rho} R^\nu{}_\rho - \frac{1}{2} R^{\rho\sigma} R_{\rho\sigma} g^{\mu\nu} + R^{\rho\sigma} g^{\mu\nu} \nabla_\rho \nabla_\sigma \\
&\hspace{6em}- 2 R^{\mu\rho} \nabla^\nu \nabla_\rho + R^{\mu\nu} \nabla^2 \Big] \delta g_{\mu\nu} \eqend{,}
\end{split} \\
\delta \left( R^2 \sqrt{-g} \right) &= - \sqrt{-g} \left[ 2 R^{\mu\nu} R - \frac{1}{2} R^2 g^{\mu\nu} - 2 R \nabla^\mu \nabla^\nu + 2 R g^{\mu\nu} \nabla^2 \right] \delta g_{\mu\nu} \eqend{,}
\end{equations}
performing integration by parts and using the second (contracted) Bianchi identities
\begin{equation}
\nabla^\mu R_{\mu\nu\rho\sigma} = 2 \nabla_{[\rho} R_{\sigma]\nu} \eqend{,} \qquad \nabla^\mu R_{\mu\nu} = \frac{1}{2} \nabla_\nu R \eqend{,}
\end{equation}
we obtain
\begin{equation}
\expect[div]{ T^{\mu\nu}(x) } = \frac{\diver}{720 (4\pi)^2} \frac{1}{\sqrt{-g}} \frac{\delta}{\delta g_{\mu\nu}(x)} \int \left( 7 R^{\mu\nu\rho\sigma} R_{\mu\nu\rho\sigma} + 8 R^{\mu\nu} R_{\mu\nu} - 5 R^2 \right) \sqrt{-g} \total^n x \eqend{,}
\end{equation}
which is indeed a local term that can be canceled by adding its negative to the action.

Consider now the difference between the divergent parts in the original~\eqref{eq:VEV1_div} and modified~\eqref{eq:VEV1_div_mod} Breitenlohner--Maison scheme\footnote{Since we will be interested in taking the trace and the $n \to 4$ limit, we already consider the integrals in four dimensions.}, which reads
\begin{splitequation}
\expect[div,mod]{ T^{\mu\nu}(x) }_{(1)} - \expect[div,orig]{ T^{\mu\nu}(x) }_{(1)} &= \frac{\diver}{120\cdot 16 \pi^2} \hat\eta^{\mu\nu} \iint \bar\Pi^{\rho\sigma}(p) p^2 \mathe^{\mathi p (x-y)} \dn[4] p h_{\rho\sigma}(y) \total^4 y \\
&= \frac{\diver}{120\cdot 16 \pi^2} \hat\eta^{\mu\nu} \left[ \nabla^2 R(x) \right]_{(1)} \eqend{,} \raisetag{1.7em}
\end{splitequation}
where we used eq.~\eqref{eq:barpi_def}, the fact that the external momentum $p$ and field $h_{\rho\sigma}$ must be taken four-dimensional, and the expansions from App.~\ref{app:geometric_expansion}. Taking the trace, we obtain using eq.~\eqref{eq:hateta_trace} that
\begin{equation}
\lim_{n \to 4} \eta_{\mu\nu} \left[ \expect[div,mod]{ T^{\mu\nu}(x) }_{(1)} - \expect[div,orig]{ T^{\mu\nu}(x) }_{(1)} \right] = - \frac{1}{60\cdot 16 \pi^2} \left[ \nabla^2 R(x) \right]_{(1)} \eqend{,}
\end{equation}
which is exactly the difference between the first-order trace anomalies~\eqref{eq:a1_godazgar_orig} and \eqref{eq:a1_godazgar_mod} computed using the definition of Godazgar and Nicolai in the original resp.\ modified Breitenlohner--Maison scheme, cf. the alternative formulation~\eqref{eq:a1_godazgar_div}.

\section{The \texorpdfstring{$h$}{h}-expansion of geometrical quantities}
\label{app:geometric_expansion}

The perturbative expansion used in this work for several geometrical quantities, up to second order in the metric perturbation $h_{\mu\nu}$, reads:
\begin{align}
\Gamma^{\rho}{}_{\mu\nu} &= \kappa \left[ \partial_{(\mu }h^{\rho }{}_{\nu )} - \frac{1}{2} \partial^{\rho }h_{\mu \nu } \right] + \kappa^2 \left[ \frac{1}{2} h^{\rho \alpha } \partial_{\alpha }h_{\mu \nu } -  h^{\rho \alpha }\partial_{(\mu }h_{\nu )\alpha } \right] + \bigo{\kappa^3} \eqend{,} \\
\begin{split}
R_{\rho\sigma\mu\nu} &= \kappa \left[ - \partial_\mu \partial_{[\rho } h_{\sigma ] \nu} + \partial_\nu \partial_{[\rho } h_{\sigma ] \mu} \right] + \kappa^2 \bigg[ - \frac{1}{2} \partial_{[\rho }h_{|\mu |}{}^{\alpha }\partial_{\sigma ]}h_{\nu \alpha } - \frac{1}{2} \partial_{\mu }h_{[\rho }{}^{\alpha }\partial_{\sigma ]}h_{\nu \alpha } \\
&\quad+ \frac{1}{2} \partial_{\nu }h_{[\rho }{}^{\alpha }\partial_{\sigma ]}h_{\mu \alpha } -  \frac{1}{2} \partial_{\mu }h_{[\rho }{}^{\alpha }\partial_{|\nu |}h_{\sigma ]\alpha } + \frac{1}{2} \partial^{\alpha }h_{\mu [\rho }\partial_{\sigma ]}h_{\nu \alpha } - \frac{1}{2} \partial^{\alpha }h_{\nu [\rho }\partial_{\sigma ]}h_{\mu \alpha } \\
&\quad+ \frac{1}{2} \partial_{\mu }h_{\alpha[\rho }\partial^\alpha h_{\sigma ]\nu } -  \frac{1}{2} \partial_{\nu }h_{\alpha[\rho }\partial^\alpha h_{\sigma ]\mu } - \frac{1}{2} \partial_{\alpha }h_{\mu [\rho }\partial^\alpha h_{\sigma ] \nu} \bigg] + \bigo{\kappa^3} \eqend{,}
\end{split} \\
\begin{split}
R_{\mu\nu} &= \kappa \left[ \partial^{\alpha }\partial_{(\mu }h_{\nu )\alpha } - \frac{1}{2} \partial^2 h_{\mu \nu } - \frac{1}{2} \partial_{\mu }\partial_{\nu }h \right] + \kappa^2 \bigg[ - \frac{1}{4} \partial_{\alpha } h \partial^{\alpha }h_{\mu \nu } + \frac{1}{2} \partial^{\alpha }h_{\mu \nu } \partial_{\beta }h_{\alpha }{}^{\beta } \\
&\quad+ \frac{1}{2} h^{\alpha \beta } \partial_{\alpha }\partial_{\beta }h_{\mu \nu } + \frac{1}{2} h^{\alpha \beta }\partial_{\mu }\partial_{\nu}h_{\alpha \beta } - \frac{1}{2} \partial^{\alpha }h_{\beta(\mu } \partial^\beta h_{\nu )\alpha } + \frac{1}{2} \partial_{\alpha }h_{(\mu }{}^{\beta }\partial^\alpha h_{\nu )\beta } \\
&\quad-  h^{\alpha \beta }\partial_{\alpha }\partial_{(\mu }h_{\nu )\beta } - \partial_{(\mu }h_{\nu )}{}^{\alpha }\partial^{\beta }h_{\alpha \beta } + \frac{1}{4} \partial_{(\mu }h^{\alpha \beta }\partial_{\nu )}h_{\alpha \beta } + \frac{1}{2} \partial_{(\mu }h_{\nu )}{}^{\alpha }\partial_{\alpha }h \bigg] + \bigo{\kappa^3} \eqend{,}
\end{split} \\
\begin{split}
R &= \kappa \left[ \partial_{\alpha } \partial_{\beta } h^{\alpha \beta } - \partial^2 h \right] + \kappa^2 \bigg[ h^{\alpha \beta } \partial^2 h_{\alpha \beta } - \frac{1}{4} \partial_{\alpha }h \partial^{\alpha }h + \partial^{\alpha }h \partial_{\beta }h_{\alpha }{}^{\beta } + h^{\alpha \beta } \partial_{\alpha } \partial_{\beta } h \\
&\quad- \partial_{\alpha }h^{\alpha \beta } \partial_{\gamma }h_{\beta }{}^{\gamma } - 2 h^{\alpha \beta } \partial_{\beta } \partial_{\gamma } h_{\alpha }{}^{\gamma } -  \frac{1}{2} \partial_{\beta }h_{\alpha \gamma } \partial^{\gamma }h^{\alpha \beta } + \frac{3}{4} \partial_{\gamma }h_{\alpha \beta } \partial^{\gamma }h^{\alpha \beta } \bigg] + \bigo{\kappa^3} \eqend{,}
\end{split} \\
R^2 &= \kappa^2 \left[ \partial_{\alpha } \partial_{\beta } h^{\alpha \beta } - \partial^2 h \right]^2 + \bigo{\kappa^3} \eqend{,} \\
\begin{split}
&\hspace{-2.5em}R^{\mu\nu} R_{\mu\nu} = \frac{1}{4} \kappa^2 \left[ 2 \partial^{\alpha }\partial_{(\mu }h_{\nu )\alpha } - \partial^2 h_{\mu \nu } - \partial_{\mu }\partial_{\nu }h \right] \left[ 2 \partial_\beta \partial^\mu h^{\nu\beta} - \partial^2 h^{\mu\nu} - \partial^\mu \partial^\nu h \right] + \bigo{\kappa^3} \eqend{,}
\end{split} \\
&\hspace{-2.5em}R_{\rho\sigma\mu\nu} R^{\rho\sigma\mu\nu} = 4 \kappa^2 \partial_\rho \partial_{[\mu} h_{\nu]\sigma} \partial^\mu \partial^{[\rho} h^{\sigma]\nu} + \bigo{\kappa^3} \eqend{.}
\end{align}
We obtained these expansions using the xPert package~\cite{xpert}.

\section{The two-argument tensorial integral \texorpdfstring{$\mathcal{I}(p,k)$}{I(p,k)}}
\label{app:integrals_2}

The explicit results for the two-argument tensorial integrals~\eqref{eq:Ipk} that we have employed in the main part of the article read:
\begin{align}
\mathcal{I}(p,k) &= \frac{\mathi}{16 \pi^2} F_{00}(k,p) \eqend{,} \\
\mathcal{I}^\mu(p,k) &= \frac{\mathi}{16 \pi^2} \left[ F_{10}(k,p) k^\mu + F_{01}(k,p) p^\mu \right] \eqend{,} \\
\begin{split}
\mathcal{I}^{\mu\nu}(p,k) &= \trlss{ \mathcal{I}^{\mu\nu}(p,k) } + \frac{1}{n} \eta^{\mu\nu} \, \mathcal{I}(p-k) \eqend{,}
\end{split} \\
\mathcal{I}^{\mu\nu\rho}(p,k) &= \trlss{ \mathcal{I}^{\mu\nu\rho}(p,k) } + \frac{3}{2 (n+2)} (p+k)^{(\mu} \eta^{\nu\rho)} \, \mathcal{I}(p-k) \eqend{,} \\
\begin{split}
\mathcal{I}^{\mu\nu\rho\sigma}(p,k) &= \trlss{ \mathcal{I}^{\mu\nu\rho\sigma}(p,k) } + \frac{3}{2 (n-1) (n+4)} \Bigl\lbrace - \left[ k^2 - \frac{6 (p\cdot k)}{n+2} + p^2 \right] \eta^{(\mu\nu} \eta^{\rho\sigma)} \\
&\hspace{3em}+ \left[ n k^{(\mu} k^\nu + 2 (n-2) k^{(\mu} p^\nu + n p^{(\mu} p^\nu \right] \eta^{\rho\sigma)} \Bigr\rbrace \, \mathcal{I}(p-k) \eqend{,} \raisetag{2em}
\end{split} \\
\begin{split}
\mathcal{I}^{\mu\nu\rho\sigma\alpha}(p,k) &= \trlss{ \mathcal{I}^{\mu\nu\rho\sigma\alpha}(p,k) } + \frac{5}{4 (n-1) (n+6)} \Bigl\lbrace \frac{6}{n+4} \left[ k^2 k^{(\alpha} + p^2 p^{(\alpha} \right] \eta^{\mu\nu} \eta^{\rho\sigma)} \\
&\qquad+ \left[ (n+2) (k+p)^{(\alpha} (k+p)^\mu - 12 k^{(\alpha} p^\mu \right] (k+p)^\nu \eta^{\rho\sigma)} \\
&\qquad- 3 \left[ k^2 - \frac{8}{n+4} (p\cdot k) + p^2 \right] (k+p)^{(\alpha} \eta^{\mu\nu} \eta^{\rho\sigma)} \Bigr\rbrace \mathcal{I}(p-k) \eqend{,} \raisetag{1.8em}
\end{split} \\
\begin{split}
I^{\mu\nu\rho\sigma\alpha\beta}(p,k) &= \trlss{ \mathcal{I}^{\mu\nu\rho\sigma\alpha\beta}(p,k) } + \frac{15}{16 (n^2-1) (n+4) (n+6) (n+8)} \eta^{(\mu\nu} \eta^{\rho\sigma} \eta^{\alpha\beta)} \\
&\qquad\times \Bigl\lbrace 3 (n+4) (n+6) \left[ (k^2)^2 + (p^2)^2 \right] - 60 (n+4) ( k^2 + p^2 ) (p\cdot k) \\
&\qquad\qquad+ 4 (n^2 + 4 n + 48) (p\cdot k)^2 + 2 (3 n^2 + 22 n + 64) k^2 p^2 \Bigl\rbrace \mathcal{I}(p-k) \\
&\quad- \frac{45 (n+2)}{8 (n^2-1) (n+6) (n+8)} \eta^{(\mu\nu} \eta^{\rho\sigma} (k+p)^\alpha (k+p)^{\beta)} \\
&\qquad\qquad\times \left[ (n+6) k^2 - 10 (p\cdot k) + (n+6) p^2 \right] \mathcal{I}(p-k) \\
&\quad+ \frac{45}{2 (n^2-1) (n+6) (n+8)} \eta^{(\mu\nu} \eta^{\rho\sigma} \Bigl\lbrace (n+1) k^\alpha k^{\beta)} p^2 + (n+1) p^\alpha p^{\beta)} k^2 \\
&\qquad\qquad+ k^\alpha p^{\beta)} \left[ (2n+7) ( k^2 + p^2 ) - 10 (p\cdot k) \right] \Bigl\rbrace \mathcal{I}(p-k) \\
&\quad+ \frac{15 (n+2)}{16 (n^2-1) (n+8)} \eta^{(\mu\nu} \Bigl\lbrace (n+4) (k+p)^\rho (k+p)^\sigma (k+p)^\alpha (k+p)^{\beta)} \\
&\qquad\qquad- 24 (k+p)^\rho (k+p)^\sigma k^\alpha p^{\beta)} + \frac{48}{n+2} k^\rho k^\sigma p^\alpha p^{\beta)} \Bigl\rbrace \mathcal{I}(p-k) \eqend{,}
\end{split}
\end{align}
where the parametric integrals $F_{ab}(p,k)$ have been defined in eq.~\eqref{eq:F}, and the traceless contributions read
\begin{equation}
\trlss{ \mathcal{I}^{\mu_1 \cdots \mu_m}(p,k) } = \sum_{j=0}^m \frac{m!}{j! (m-j)!} F_{j,m-j}(k,p) \trlss{ k^{(\mu_1} \cdots k^{\mu_j} p^{\mu_{j+1}} \cdots p^{\mu_m)} } \eqend{.}
\end{equation}
The traceless part on the right-hand side can be computed from formula~\eqref{eq:traceless}, taking $\alpha k + \beta p$ in that formula, performing the appropriate number of derivatives with respect to $\alpha$ and $\beta$, and setting $\alpha = \beta = 0$ in the result.

We note that a consistency check can be made on these results. Namely, from the definition~\eqref{eq:Ipk} of the two-argument tensorial integral one computes analogously to the trace condition~\eqref{eq:Ipk_trace} that
\begin{splitequation}
\label{eq:Ipk_pcontraction}
2 p_\mu \, \mathcal{I}^{\mu\nu \cdots}(p,k) &= p^2 \, \mathcal{I}^{\nu \cdots}(p,k) - \int \frac{q^\nu \cdots}{q^2 (q-k)^2} \dn q + \int \frac{q^\nu \cdots}{(q-p)^2 (q-k)^2} \dn q \\
&= p^2 \, \mathcal{I}^{\nu \cdots}(p,k) - \mathcal{I}^{\nu \cdots}(-k) + \mathcal{I}^{\nu \cdots}(p-k) + p^\nu \, \mathcal{I}^{\cdots}(p-k) + \cdots \eqend{,}
\end{splitequation}
where we have shifted the integration variable $q \to q+p$ in the last integral. Using the recursion relations for the $F_{ab}$ given in App.~\ref{app:FG}, Eq.~\eqref{eq:Ipk_pcontraction} can be shown to hold after a long but straightforward computation.

\section{The parameter integrals \texorpdfstring{$F_{ab}$}{Fab}}
\label{app:FG}

Here we prove relations for the parameter integrals $F_{ab}$ defined in eq.~\eqref{eq:F}, which can be used for a consistency check on the results for the two-argument tensorial integral~\eqref{eq:Ipk} given in App.~\ref{app:integrals_2}. We recall the definitions~\eqref{eq:F} and~\eqref{eq:me_def}:
\begin{equation}
F_{ab}(k,p) = \int_0^1 \int_0^{1-y} \frac{x^a y^b}{x (1-x) k^2 + y (1-y) p^2 - 2 x y (p\cdot k) - \mathi 0} \total x \total y \eqend{,}
\end{equation}
and, for better readability, leave out the arguments $(k,p)$ of $F_{ab}$ in the remainder of this appendix. For $a,b > 0$ it follows directly that
\begin{splitequation}
\label{eq:appFG_pkf_1}
&k^2 F_{a,b-1} - k^2 F_{a+1,b-1} + p^2 F_{a-1,b} - p^2 F_{a-1,b+1} - 2 (p\cdot k) F_{ab} \\
&\quad= \int_0^1 \int_0^{1-y} x^{a-1} y^{b-1} \total x \total y = \frac{(a-1)! (b-1)!}{(a+b)!} \eqend{,}
\end{splitequation}
which can be used to replace $(p\cdot k) F_{ab}$. If instead $b = 0$ and $a > 0$, we compute
\begin{splitequation}
\label{eq:appFG_pkf_2}
&2 (p\cdot k) F_{a0} - p^2 F_{a-1,0} + 2 p^2 F_{a-1,1} \\
&\quad= \int_0^1 \int_0^{1-y} \frac{x^{a-1} [ 2 x (p\cdot k) - (1-2y) p^2 ]}{x (1-x) k^2 + y (1-y) p^2 - 2 x y (p\cdot k) - \mathi 0} \total x \total y \\
&\quad= - \int_0^1 \int_0^{1-x} x^{a-1} \partial_y \ln \left[ x (1-x) k^2 + y (1-y) p^2 - 2 x y (p\cdot k) - \mathi 0 \right] \total y \total x \\
&\quad= - \int_0^1 x^{a-1} \total x \ln \left[ \frac{(k-p)^2 - \mathi 0}{k^2 - \mathi 0} \right] = - \frac{1}{a} \ln \left[ \frac{(k-p)^2 - \mathi 0}{k^2 - \mathi 0} \right] \eqend{,}
\end{splitequation}
and analogously for $a = 0$ and $b > 0$
\begin{equation}
\label{eq:appFG_pkf_3}
(p\cdot k) F_{0b} = \frac{1}{2} k^2 F_{0,b-1} - k^2 F_{1,b-1} - \frac{1}{2 b} \ln \left[ \frac{(k-p)^2 - \mathi 0}{p^2 - \mathi 0} \right] \eqend{,}
\end{equation}
such that we can replace all $(p\cdot k) F_{ab}$ with $a,b \geq 0$.

Furthermore, we compute for $a,b \geq 0$ that
\begin{splitequation}
\label{eq:appF_gla}
&2 k^2 F_{a+2,b} = - \int_0^1 \int_0^{1-y} x^{a+1} y^b \partial_x \ln \left[ x (1-x) k^2 + y (1-y) p^2 - 2 x y (p\cdot k) - \mathi 0 \right] \total x \total y \\
&\hspace{5em}- 2 (p\cdot k) F_{a+1,b+1} + k^2 F_{a+1,b} \\
&\quad= - \int_0^1 (1-y)^{a+1} y^b \left( \ln \left[ y (1-y) \right] + \ln \left[ (k-p)^2 - \mathi 0 \right] \right) \total y \\
&\hspace{2em}+ (a+1) \int_0^1 \int_0^{1-y} x^a y^b \ln \left[ x (1-x) k^2 + y (1-y) p^2 - 2 x y (p\cdot k) - \mathi 0 \right] \total x \total y \\
&\hspace{2em}- 2 (p\cdot k) F_{a+1,b+1} + k^2 F_{a+1,b} \eqend{,} \raisetag{1.5em}
\end{splitequation}
where we integrated by parts in $x$, and analogously
\begin{splitequation}
\label{eq:appF_glb}
&2 p^2 F_{a,b+2} = - \int_0^1 x^a (1-x)^{b+1} \left( \ln \left[ x (1-x) \right] + \ln \left[ (k-p)^2 - \mathi 0 \right] \right) \total x \\
&\qquad+ (b+1) \int_0^1 \int_0^{1-y} x^a y^b \ln \left[ x (1-x) k^2 + y (1-y) p^2 - 2 x y (p\cdot k) - \mathi 0 \right] \total x \total y \\
&\qquad- 2 (p\cdot k) F_{a+1,b+1} + p^2 F_{a,b+1} \eqend{.} \raisetag{1.5em}
\end{splitequation}
Performing the integrals over $y$ or $x$ in the first line of the right-hand sides of eqs.~\eqref{eq:appF_gla} and~\eqref{eq:appF_glb} and equating the double integrals in the second line, it follows that
\begin{splitequation}
(a+1) p^2 \left( 2 F_{a,b+2} - F_{a,b+1} \right) &= (b+1) k^2 \left( 2 F_{a+2,b} - F_{a+1,b} \right) \\
&\quad+ (b-a) \left[ \frac{a! b!}{(a+b+2)!} + 2 (p\cdot k) F_{a+1,b+1} \right] \eqend{,}
\end{splitequation}
and using the relation~\eqref{eq:appFG_pkf_1} we obtain
\begin{equation}
\label{eq:appF_k2p2f}
(a+b+2) k^2 F_{a+2,b} - (a+1) k^2 F_{a+1,b} = (a+b+2) p^2 F_{a,b+2} - (b+1) p^2 F_{a,b+1} \eqend{.}
\end{equation}

\clearpage
\addcontentsline{toc}{section}{References}
\bibliography{traceanom}
  
\end{document}